\documentclass[%
 aip,
 amsmath,amssymb,
 reprint,%
]{revtex4-1}

\usepackage{bibentry}

\usepackage{graphicx}
\usepackage{dcolumn}
\usepackage{bm}

\usepackage[utf8]{inputenc}
\usepackage[T1]{fontenc}
\usepackage{mathptmx}
\usepackage{etoolbox}

\makeatletter
\def\@email#1#2{%
 \endgroup
 \patchcmd{\titleblock@produce}
  {\frontmatter@RRAPformat}
  {\frontmatter@RRAPformat{\produce@RRAP{*#1\href{mailto:#2}{#2}}}\frontmatter@RRAPformat}
  {}{}
}%
\makeatother
\begin{document}

\preprint{AIP/123-QED}

\title{Free-electron-light interactions in nanophotonics}
\author{Charles Roques-Carmes$^{*}$}
\affiliation{Research Laboratory of Electronics, MIT, Cambridge, MA 02139, USA}%
\email{chrc@mit.edu}

\author{Steven E. Kooi}
\affiliation{Institute for Soldier Nanotechnologies, MIT, Cambridge, MA, 02139, USA}%

\author{Yi Yang}
\affiliation{Research Laboratory of Electronics, MIT, Cambridge, MA 02139, USA}%
\affiliation{Department of Physics, MIT, Cambridge, MA 02139, USA}%
\affiliation{Department of Physics, University of Hong Kong, Hong Kong, China}

\author{Nicholas~Rivera}
\affiliation{Department of Physics, MIT, Cambridge, MA 02139, USA}%

\author{Phillip~D.~Keathley}
\affiliation{Research Laboratory of Electronics, MIT, Cambridge, MA 02139, USA}%

\author{John~D.~Joannopoulos}
\affiliation{Institute for Soldier Nanotechnologies, MIT, Cambridge, MA, 02139, USA}%
\affiliation{Department of Physics, MIT, Cambridge, MA 02139, USA}%

\author{Steven~G.~Johnson}
\affiliation{Department of Mathematics, MIT, Cambridge, MA, 02139, USA}

\author{Ido~Kaminer}
\affiliation{Department of Electrical and Computer Engineering, Technion Haifa 32000, Israel}

\author{Karl K. Berggren}
\affiliation{Research Laboratory of Electronics, MIT, Cambridge, MA 02139, USA}%

\author{Marin Solja\v{c}i\'{c}}
\affiliation{Research Laboratory of Electronics, MIT, Cambridge, MA 02139, USA}%
\affiliation{Department of Physics, MIT, Cambridge, MA 02139, USA}%


\date{\today}

\begin{abstract}
When impinging on optical structures or passing in their vicinity, free electrons can spontaneously emit electromagnetic radiation, a phenomenon generally known as cathodoluminescence. Free-electron radiation comes in many guises: Cherenkov, transition, and Smith-Purcell radiation, but also electron scintillation, commonly referred to as incoherent cathodoluminescence. While those effects have been at the heart of many fundamental discoveries and technological developments in high-energy physics in the past century, their recent demonstration in photonic and nanophotonic systems has attracted a lot of attention. Those developments arose from predictions that exploit nanophotonics for novel radiation regimes, now becoming accessible thanks to advances in nanofabrication. In general, the proper design of nanophotonic structures can enable shaping, control, and enhancement of free-electron radiation, for any of the above-mentioned effects. Free-electron radiation in nanophotonics opens the way to promising applications, such as widely-tunable integrated light sources from x-ray to THz frequencies, miniaturized particle accelerators, and highly sensitive high-energy particle detectors. Here, we review the emerging field of \emph{free-electron radiation in nanophotonics}. We first present a general, unified framework to describe free-electron light-matter interaction in arbitrary nanophotonic systems. We then show how this framework sheds light on the physical underpinnings of many methods in the field used to control and enhance free-electron radiation. Namely, the framework points to the central role played by the photonic eigenmodes in controlling the output properties of free-electron radiation (e.g., frequency, directionality, and polarization). We then review experimental techniques to characterize free-electron radiation in scanning and transmission electron microscopes, which have emerged as the central platforms for experimental realization of the phenomena described in this Review. We further discuss various experimental methods to control and extract spectral, angular, and polarization-resolved information on free-electron radiation. We conclude this Review by outlining novel directions for this field, including ultrafast and quantum effects in free-electron radiation, tunable short-wavelength emitters in the ultraviolet and soft x-ray regimes, and free-electron radiation from topological states in photonic crystals.

\end{abstract}

\maketitle

\tableofcontents

\section{Introduction}


The interaction of free electrons with light and matter is a century-old field of research, that has had profound implications in electron microscopy, radiation sources, and high-energy particle detection. At the heart of this field lies a few fundamental discoveries, unveiling various conditions in which free electrons can convert part of their energy into photons.~\cite{Crooke1879XVI.Discharge, Cherenkov1934VisibleRadiation, Ginzburg1945, Smith1953, Ritchie1957PlasmaFilms} On the other hand, over the past two decades, nanophotonics has emerged as a platform to control photonic modes at the nanoscale, by patterning materials at scales comparable to the photon wavelength. The recent merger of these two fields has spurred new applications and fundamental discoveries.\cite{Luo2003CerenkovCrystals, Bashevoy2006GenerationImpact, VanWijngaarden2006DirectSpectroscopy, Bashevoy2007HyperspectralResolution, Vesseur2007DirectSpectroscopy, Hofmann2007PlasmonicCathodoluminescence, Adamo2009LightChip, Adamo2010TuneableSource, Barnard2011ImagingCathodoluminescence, Adamo2012, Kaminer2017, Liu2017IntegratedThreshold, Massuda2017, Yang2018MaximalElectrons, Sapra2020On-chipAccelerator, Shentcis2020TunableMaterials, Kfir2020ControllingModes, Wang2020CoherentCavity, Nussupbekov2021EnhancedNanowell, Roques-Carmes2022ANanophotonics, GarciaDeAbajo2010OpticalMicroscopy, Polman2019Electron-beamNanophotonics, Talebi2017InteractionFunction, Coenen2016AGeology, Christopher2020Electron-drivenMicroscopes} This Review is dedicated to (1) providing a general understanding of free-electron-light interactions mediated by nanophotonic structures; (2) highlighting recent theoretical and experimental developments in the field; (3) outlining future prospects for fundamental research and novel applications.


How can charged particles emit light? This fundamental question has driven much development in theoretical and experimental physics in the twentieth century. Perhaps the original interest in this question can be traced back to early discoveries in radioactivity, where luminescence from liquids was used to detect the presence of radioactive substances.~\cite{MarieCurie1904Radio-activeSubstances, Ginzburg1996RadiationPhenomena} Later on, emission processes such as the Cherenkov effect were extensively used to track and detect particles.~\cite{Akopov2002TheDetector, Adams2001TheDetector,Nakamura2003Hyper-KamiokandeDetector, Kleinknecht1982ParticleDetectors} Therefore, the understanding of radiation processes from charged particles, such as free electrons, has evolved in tandem with some of the most profound discoveries of modern physics, such as quantum electrodynamics and particle detection within and beyond the standard model.~\cite{Nakamura2003Hyper-KamiokandeDetector, Papanestis2014TheOperation} Concepts from free-electron radiation have also permeated throughout many fields of physics, from nonlinear optics to gravitational physics.~\cite{Carusotto2013}  Other forms of spontaneous emission induced by free-electron (de)acceleration -- which are not covered in this Review -- have received a lot of interest, especially in the context of free-electron lasers.~\cite{Pellegrini2016TheLasers}



Much more recently, nanophotonics has become a paramount framework and technology, enabling, among other things, the design of novel light sources, detectors, and devices controlling the polarization, spectral, and angular distribution of light.~\cite{Joannopoulos2011,Novotny2009PrinciplesNano-optics} A hallmark of nanophotonics is the design of nanostructured materials (metasurfaces,~\cite{Yu2014FlatMetasurfaces} photonic crystals,~\cite{Joannopoulos2011,Yablonovitch1994PhotonicCrystals} resonators,~\cite{Vahala2003OpticalMicrocavities, Armani2003Ultra-high-QChip} etc.) to tailor the interaction of light with matter, either by shaping light propagation at the nanoscale, or by controlling emission from atoms and molecules. 

Free electrons and other types of charged high-energy particles usually carry large kinetic energies compared to the energies of optical photons often controlled with nanophotonics, and can in principle emit photons with any energy below the kinetic energy of the electron (including at optical frequencies). The perspective of enhancing and controlling free-electron radiation with nanophotonics thus applies to wide spectral ranges.


There has been a recent surge of interest in research at the intersection of free-electron physics and nanophotonics.~\cite{GarciaDeAbajo2010OpticalMicroscopy, Polman2019Electron-beamNanophotonics} If this research is successful, nanophotonics-enhanced free-electron light sources could cover the entire electromagnetic spectrum, with controllable polarization, spectral, spatial, and angular properties. This perspective is all the more attractive for regions of the electromagnetic spectrum where sources are scarce, inefficient, bulky, and/or expensive (such as THz, deep ultraviolet (UV), and x-rays), enabling novel lab-on-chip applications. Unveiling novel regimes of free-electron radiation in nanophotonic systems would also open the way to enhanced beam diagnosis and detection tools, such as Cherenkov and scintillation detectors which are ubiquitous in many domains of modern science and engineering.~\cite{Akopov2002TheDetector, Adams2001TheDetector, Nakamura2003Hyper-KamiokandeDetector, Kleinknecht1982ParticleDetectors, Gektin2017InorganicSystems, Shaffer2017} Such detectors could, for instance, leverage various effects in nanophotonics to significantly increase their sensitivity and strongly discriminate signals from various incident particles.~\cite{Lin2021ADetectors,Lin2018ControllingRadiation} 

The inverse effects have also attracted a great amount of attention: in nanostructures absorbing energy from powerful lasers, particles can be accelerated, forming the basis for highly compact particle accelerators that may even one day fit on a chip.~\cite{Niedermayer2017DesigningChip, Black2019Laser-DrivenMicrostructures, Sapra2020On-chipAccelerator} Nanophotonic particle accelerators exhibit much higher damage threshold and acceleration gradient than conventional linear accelerators and their compact form factor opens the perspectives of point-of-care radiation medicine and table-top high-energy electron microscopes. One should expect other areas of nanophotonics, such as topological photonics and design optimization (via inverse design and topology optimization), to have an equivalently important impact on nanophotonics-enhanced free-electron physics. In this context, we wish to provide a unified picture of free-electron radiation in nanophotonic structures, highlighting physical processes to control and enhance radiation, thereby enabling some of the applications mentioned above. 

In all of the above-mentioned applications, recent works~\cite{Karnieli2021TheParticle,BenHayun2021ShapingElectrons,Wong2021ControlWavepackets,GarciaDeAbajo2021OpticalOpportunities,DiGiulio2021ModulationLight,Kaminer2016QuantumMomentum,Kfir2021OpticalElectrons,Rivera2020LightmatterQuasiparticles, Adiv2021QuantumAccelerators} have highlighted the possibility of shaping the quantum properties of free electrons and the emitted radiation. This branch of work could also open the way to novel sources of quantum light with controllable properties. 

Previous works reviewed experimental results in electron-light interactions, providing frameworks to calculate electron energy loss spectroscopy (EELS) and cathodoluminescence (CL) in classical and quantum regimes.~\cite{GarciaDeAbajo2010OpticalMicroscopy} More recent reviews highlighted spectroscopy techniques combining the unprecedented combination of high space, energy, and time resolution enabled by electron beams, with a focus on quantum and ultrafast effects.~\cite{Polman2019Electron-beamNanophotonics}

In this Review, we highlight the role of nanophotonics in free-electron physics and electron-light interactions. We show how one can control and enhance the interaction of electron beams with photonic modes for various types of free-electron radiation physics. We first give a high-level overview of several types of free-electron radiation processes, followed by a historical timeline of the field of free-electron physics, and an outline of some of the recent achievements enabled in this field by nanophotonics. In section~\ref{sec:typology}, we then revisit the typology of free-electron radiation with a general formalism accounting for most types of coherent (sections~\ref{sec:theory-cherenkov}, \ref{sec:theory-sp}, \ref{sec:theory-tr}, and \ref{sec:theory-spp}) and incoherent cathodoluminescence (section~\ref{sec:theory-scint}). With building blocks of the formalism outlined in section~\ref{sec:theory}, we revisit several types of free-electron radiation as a form of interaction between a free electron and specific photonic eigenmodes in a nanophotonic structure. We also connect our formalism to recent works on calculating bounds for free-electron radiation and energy loss in nanophotonics (section~\ref{sec:theory-bounds}). In section~\ref{sec:expt}, we review experimental methods and considerations to observe and quantify such effects (section~\ref{sec:expt-details}), and describe nanophotonic techniques to control (section~\ref{sec:control}) and enhance (section~\ref{sec:enhance}) (coherent and incoherent) cathodoluminescence. We then outline several exciting perspectives at the intersection of nanophotonics and free-electron physics in section~\ref{sec:future}. We conclude this Review in section~\ref{sec:conclusion} by summarizing our main findings, progress in the field, and future applications of this active field of research.

\subsection{Free-electron-light interaction mediated by nanophotonic structures}
\label{intro-pm}

Free electrons can emit light in many different ways. Radiation generally occurs when the electron (with charge $q$ and propagating at velocity $\mathbf{v}$) interacts with a structure or medium supporting photonic modes such that energy-momentum conservation is satisfied. In general, one can predict which modes are excited by a free electron by considering the phase-matching condition:~\cite{Schachter1997}
\begin{equation}
    \label{eq:phase-matching}
    \omega = \mathbf{v}\cdot\mathbf{k},
\end{equation}
where $\omega$ is the photon frequency, and $\mathbf{k}$ its wavevector. This condition requires that the electron velocity and the mode phase velocity coincide. From this formula, many situations in free-electron radiation can be readily analyzed. For example, one immediate consequence of Eq.~(\ref{eq:phase-matching}) is that in free space, free-electron radiation from uniformly moving particles is prohibited. It would require the electron to move at the phase velocity of light in vacuum. Unless stated otherwise, in the following, we will consider \textit{point} electrons propagating in rectilinear motion defined by $\mathbf{v}$ in three-dimensional space. This is in contrast to \textit{sheet} electrons, which can be considered in two-dimensional problems as a mathematical convenience that reduces computational complexity. In the latter case, the transverse component of the momentum in Eq.~(\ref{eq:phase-matching}) can be neglected.  

Nanophotonic structures with various geometries and symmetries enable the control of the dispersion relation $\omega(\mathbf{k})$. In particular, because of the ability of these structures to reduce the phase velocity of light, they enable radiation in situations where unstructured materials may not. In periodic media, Bloch modes generally have an infinite Fourier series of components at wavevectors given by $\mathbf{k} = \mathbf{k'} + \mathbf{G}$, where $\mathbf{k'}$ lies in the first Brillouin Zone (BZ) and $\mathbf{G}$ is a reciprocal lattice vector. This means that Eq.~(\ref{eq:phase-matching}) can be satisfied at arbitrarily small velocities $\mathbf{v}$, which is why Smith-Purcell radiation (SPR) -- an effect we discuss extensively in this Review -- has no low-velocity cutoff. Another important consequence of Eq.~(\ref{eq:phase-matching}) is that for structures that break translation symmetry (e.g., a photonic crystal defect cavity or a plasmonic nanoparticle), photonic modes have all possible wavevectors, allowing photon emission into any localized mode. Another example in which translation-symmetry-breaking is important is in transition radiation, discussed below, where the interface between two materials is chiefly responsible for the emitted light.

\begin{figure*}
\includegraphics[scale = 0.8]{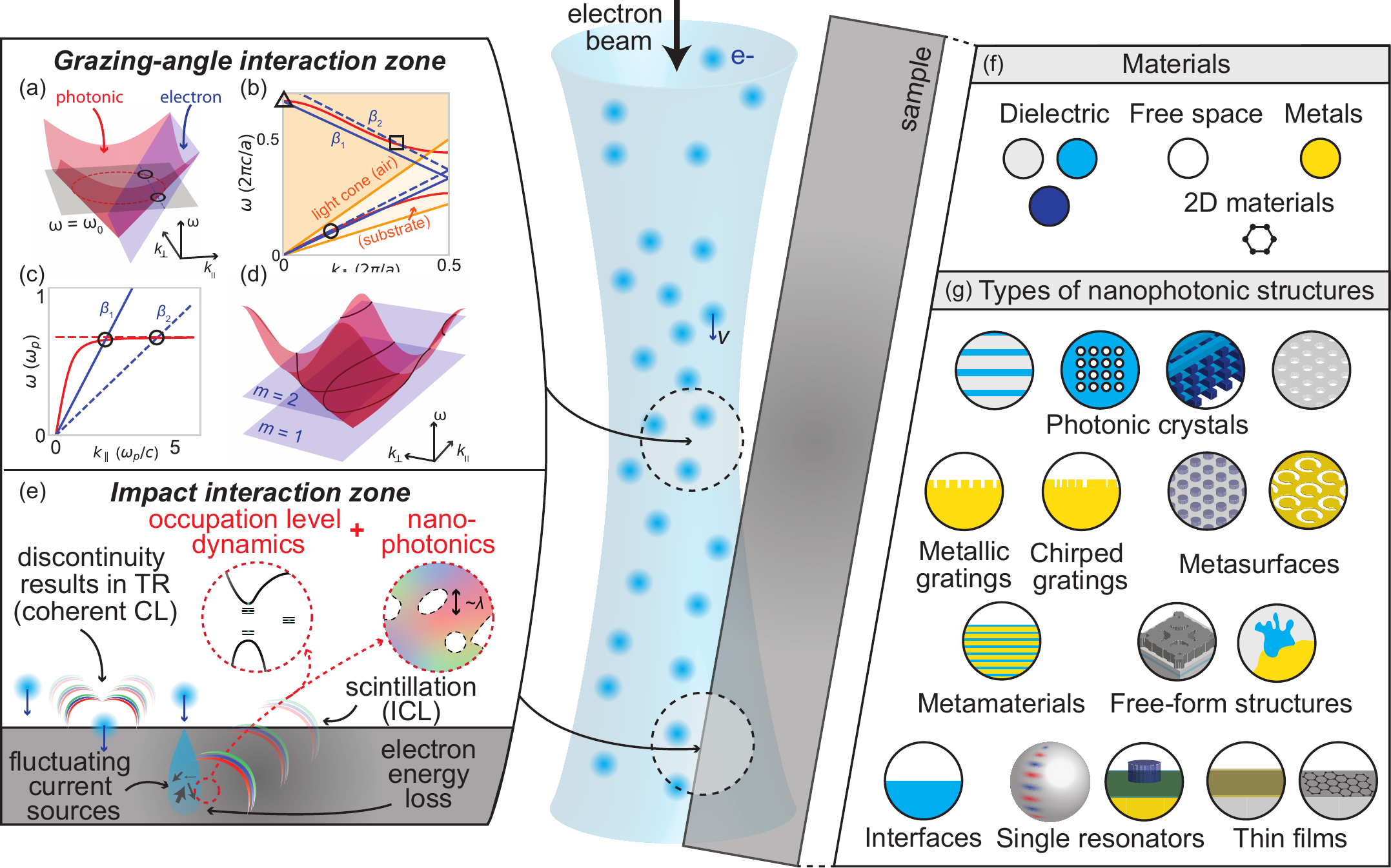}
\caption{\label{fig:overview} \textbf{Overview of free-electron radiation in nanophotonics.} Schematic of an electron beam interacting with a sample. Radiation can occur from events in the "grazing-angle interaction zone" (a-d) or in the "impact interaction zone" (e). In this Review, we discuss interactions of free-electrons with nanophotonic structures made of materials as listed in (f), and of geometries as listed in (g). In (a-d), the electrons' surfaces/lines are shown in blue and photonic mode band structures are shown in red. (a) Schematic band structure of Cherenkov radiation in uniform medium. The black plane represents an isofrequency plane with intersections of red and blue surfaces shown as dashed lines. (b) Schematic band structure of free electrons interacting with a one-dimensional periodic photonic crystal (PhC) slab waveguide. Free electrons propagating at velocity $\beta_1$ (full line) can excite guided modes (circles) and symmetry-protected bound states in the continuum (triangle). Free electrons propagating at velocity $\beta_2$ (dashed line) can excite guided modes (circles) and leaky resonances (square). (c) Dispersion plot of plasmon polariton (SPP) excitation by free electrons flying above a vacuum-metal interface. Electrons propagating at various speeds can excite modes of various energies under the cut-off ($\omega_p/\sqrt{2}$). (d) Dispersion plot of free-electron radiation in a two-dimensional periodic PhC. Electron planes with various radiation orders $m\in\mathbb{N}$ can intersect the photonic band structure. (e) In the impact interaction zone, emission can occur at the boundary between vacuum and sample (resulting in transition radiation (TR)); or from emitters in the sample excited in the electron energy loss volume (shown in blue). Emission from fluctuating current sources can be modeled by taking into account their occupation level dynamics and the nanophotonic environment in which they emit.~\cite{Kurman2020Photonic-CrystalDetection, Roques-Carmes2022ANanophotonics}}
\end{figure*}

A typical interaction of an electron beam with a sample is shown in Fig.~\ref{fig:overview}. A beam of electrons interacts with a sample in two distinct manners, corresponding to the grazing-angle and impact interaction zones. We now proceed to discuss each of these two zones separately.

\subsubsection{Grazing-angle interaction zone}

In the grazing-angle interaction zone, the phase-matching condition from Eq.~(\ref{eq:phase-matching}) allows us to predict which photonic modes are excited by free electrons, given prior knowledge of the photonic modes, as shown in Fig.~\ref{fig:overview}(a-d). This general principle accounts for the excitation of localized and extended modes, in various photonic environments. 

More specifically, as shown in Fig.~\ref{fig:overview}(a-d), electrons passing in the vicinity of photonic structures can transfer some of their energy into available modes. Examples of such interactions include Cherenkov radiation from free electrons propagating in a uniform medium (Fig.~\ref{fig:overview}(a)); excitation of guided, radiative, and bound states in the continuum (BIC) modes in photonic crystals (PhC) (Fig.~\ref{fig:overview}(b)); excitation of surface plasmon polaritons (SPP) localized at metal-dielectric interfaces (Fig.~\ref{fig:overview}(c)); and excitation of a superposition of Bloch modes in a two-dimensional PhC (Fig.~\ref{fig:overview}(d)).

In a bulk medium, free electrons propagating faster than the phase velocity of light in the medium can excite photonic modes (such that the velocity component of the photon phase velocity is equal to the electron velocity, as per Eq.~(\ref{eq:phase-matching})). In a non-dispersive dielectric, such an electron thereby emits Cherenkov radiation, a "shock wave" of light.\cite{Cherenkov1934VisibleRadiation,Ginzburg1940QuantumMedium,Ginzburg1996RadiationPhenomena,Cox1944MomentumRadiation} In Fig.~\ref{fig:overview}(a), the free-electron plane (defined as $\omega=|\mathbf{v}|k_\parallel$) intersects the light cone (defining bulk medium plane waves) only above a certain threshold velocity $\beta = 1/n$ (where $\beta = v/c$ is the reduced velocity). At a given frequency (isofrequency plane shown in black in Fig.~\ref{fig:overview}(a)), the intersection of the two surfaces (consisting of two points at $\pm k_\perp$) defines the angle of emission (see section~\ref{sec:theory-cherenkov} for more details). 

When propagating in periodic media, free electrons can spontaneously emit photons in the form of Bloch modes.\cite{Smith1953} The folded free-electron line (for sheet electrons) or plane (for point electrons) can intersect photonic bands at various locations in the BZ. Two such scenarios are shown in Fig.~\ref{fig:overview}(b,d) for the case of a one-dimensional PhC slab (interacting with a sheet electron) and a two-dimensional PhC (interacting with a point electron). In both cases, free electrons can excite modes from several bands at the same time, resulting in complex emission processes. In PhC slabs, guided modes and leaky resonances can be excited, and even modes with diverging quality factors (so-called bound states in the continuum (BIC)\cite{Hsu2013ObservationContinuum}). Emission patterns in the two-dimensional case can exhibit even richer physics, given that radiation arises as a superposition of multiple modes with various group velocities.\cite{Luo2003CerenkovCrystals}

Other types of spatially-extended modes, such as SPPs, phonon polaritons, and other surface waves can be excited by free electrons flying in their vicinity (for SPPs, typically, parallel to the interface between a metal and a dielectric). The excitation of SPPs with free electrons was originally proposed by Ritchie.~\cite{Ritchie1957PlasmaFilms} The typical dispersion relation converging asymptotically to $\omega_p/\sqrt{2}$ (where $\omega_p$ is the plasma frequency) intersects the electron dispersion at various ($\omega, k_\parallel$) as a function of the electron velocity. Such as dispersion relation is shown in Fig.~\ref{fig:overview}(c) for the case of a sheet electron. Similar analysis can be performed to account for the excitation of high-$Q$ resonances in, e.g. optical beads exhibiting whispering-gallery modes.\cite{Kfir2020ControllingModes, Muller2021BroadbandResonators}

\subsubsection{Impact interaction zone}

In the impact interaction zone, one way in which electrons can radiate is transition radiation (TR), which occurs when an electron crosses the interface between two media with distinct electromagnetic properties. From an electromagnetic perspective, TR originates in the continuity relation of the fields at the interface.~\cite{GarciaDeAbajo2010OpticalMicroscopy} This effect might still be explained in terms of the excitation of photonic modes by a free-electron current source, using the image charge formalism (see section~\ref{sec:theory-tr}). 


Beyond the simple TR effect, the more general family of impact interaction effects appear in almost all experiments, whenever some fraction of the incoming electrons penetrates into the sample.
In this \emph{impact interaction zone}, the electron kinetic energy can be transferred to the material, allowing subsequent emission processes. 

Those emission processes can be understood as arising from fluctuating current sources associated with bound polarization (e.g., from excited electrons in defects, excitons, etc.). Such bound polarization is qualitatively similar to electrons in atoms or molecules, and thus their light emission (and nanophotonic shaping perspective) is understood from the perspective of bound-electron radiation engineering, which is a dominant paradigm in nanophotonics currently. Therefore, in this process, the phase-matching condition does not describe the coupling of free electrons with light. However, photonic engineering can still be used to enhance and control radiation.

This type of radiation is commonly referred to as incoherent cathodoluminescence (ICL). Since this process is equivalent to what is referred to as scintillation in other fields of physics (e.g. as observed with x-rays, $\gamma$-rays,~\cite{Hall2012RadiobiologyRadiologist} $\alpha$ and $\beta$ particles~\cite{Birks1967TheInstrumentation}), we will refer to this effect as "electron scintillation" as well, to highlight their common physical origin and similarity. Scintillation is a complex ``multi-physics'' process spanning several disparate length, time, and energy scales. The key steps in scintillation can be summarized as follows: (1) ionization of electrons in the sample by the pump electrons followed by production and diffusion of secondary electrons \cite{Klein1968BandgapSemiconductors}; (2) establishment of a non-equilibrium steady-state of bound electrons \cite{Wurfel1982TheRadiation,Greffet2018LightLaw}; and (3) recombination, leading to light emission (when the recombination is radiative). The final step of light emission is particularly complex in nanophotonics environments, as it results from fluctuating, spatially-distributed dipoles with a non-equilibrium distribution function. The final step also hints at the feasibility of enhancing and controlling ICL with nanophotonics. Moreover, such ICL is often ubiquitously present in experimental studies of CL, and much analysis is typically devoted to attributing signals to CL versus ICL/scintillation.~\cite{Brenny2014QuantifyingMetals}


We note that there exist additional free-electron radiation processes such as bremsstrahlung ("braking" radiation) that arises from abrupt deceleration. We do not focus on these processes in this Review because their enhancement or shaping with nanophotonics structures has not been (yet) demonstrated. In contrast, free-electron radiation effects such as parametric X-ray and coherent bremsstrahlung are discussed below to highlight their connections to the wider family of diffraction radiation phenomena that can be affected by nanophotonics. We do however discuss connections between the previously-mentioned acceleration/deceleration effects (in both interaction zones) and other forms of "diffraction radiation"\cite{Potylitsyn2010} such as parametric x-ray emission and coherent bremsstrahlung. 

\subsection{Fundamental discoveries in free-electron physics}

The fundamental physics of radiation by free-electrons (without nanophotonics) has been known for decades in the context of macroscopic electrodynamics and high-energy physics. The fundamental physics of electron-light interaction has a long history, dating back to the early 1900's and has resulted in several cornerstone discoveries in fundamental physics. 

Perhaps the most celebrated of those effects is Cherenkov radiation (CR), given its many analogues in other systems,\cite{Carusotto2013} applications in nonlinear optics,~\cite{Yuan2011HighlyFiber, Chang2010HighlyGeneration, Skryabin2003SolitonFibers, Zhang2013EnhancedFiber, Belkin2015} high-energy particle detectors,~\cite{Nakamura2003Hyper-KamiokandeDetector} dosimetry, medical imaging and therapy.\cite{Shaffer2017} The original observation of CR was reported by Cherenkov\cite{Cherenkov1934VisibleRadiation} and Vavilov\cite{Vavilov1934} from secondary (Compton) electrons in a liquid irradiated by $\gamma$-rays. Shortly thereafter, the observation was confirmed by a series of observations and theoretical predictions by Cherenkov, Vavilov, Frank, and Tamm.\cite{Frank1937CoherentMatter, Cherenkov1937TheMedium, Cherenkov1938SpatialElectrons} Cherenkov, Frank, and Tamm were awarded the Nobel Prize in Physics in 1958 for the "discovery and the interpretation of the Cherenkov effect" (a few years after Vavilov had passed away). Tamm insisted in his Nobel lecture\cite{IETamm1959GeneralPhysics} that the effect should rather be named the "Vavilov-Cherenkov effect", to highlight the contribution of Vavilov. Ginzburg later noted (regretfully) that the name "Vavilov" had been dropped in most instances.\cite{Ginzburg1996RadiationPhenomena} CR is discussed in greater depth in section~\ref{sec:theory-cherenkov}.

TR was originally proposed by Ginzburg and Tamm in 1945.\cite{Ginzburg1945,Ginzburg1946ToRadiation} The original observation was reported by Goldsmith and Jelly in 1959 in the visible by bombarding metallic surfaces with 1~MeV protons. Significant important contributions to the field of CR and TR were reported in the few decades following their original discovery, such as TR calculations from metallic thin films\cite{Garibyan1960TransitionLosses} discovery of the anomalous Doppler effect in the Cherenkov cone,\cite{Frank1942DopplerMedium,Ginzburg1947OnVelocity, Lin2018ControllingRadiation} and quantum recoil corrections to the Cherenkov effect.~\cite{Ginzburg1940QuantumMedium,Cox1944MomentumRadiation, Kaminer2016QuantumMomentum} Both techniques became mainstay technologies in high-energy particle detector experiments.\cite{Akopov2002TheDetector, Adams2001TheDetector, Nakamura2003Hyper-KamiokandeDetector, Kleinknecht1982ParticleDetectors} 
Free-electron injection into a metal can also lead to the generation of SPPs. This fundamental discovery was first proposed by Ritchie in 1957~\cite{Ritchie1957PlasmaFilms} and experimentally observed for the first time in 2006.~\cite{Bashevoy2006GenerationImpact} 

Scintillation (from various high-energy particles, such as x-rays, free electrons, and $\alpha$-particles) was originally discovered as a diagnosis and detection tool in early works on gemology and radioactivity. Early works by Hittorf (reported in Ref.\cite{Urbain1909LaRares}) and later Crookes\cite{Crooke1879XVI.Discharge} reported ICL from various stones, including diamonds. The first reported scintillator detector was invented by Crookes in 1903 to detect $\alpha$-particles, following original observations of light emission from phosphorescent powders in cathode-ray tubes in 1879.\cite{Crooke1879XVI.Discharge} Developments in optical amplification devices made scintillator detectors widely available for applications in radiology, electron microscopy, and high-energy particle detection.\cite{Gektin2017InorganicSystems} Interestingly, the first x-ray images following R\"{o}ntgen's discovery were not performed with scintillators, but rather radiation-sensitive photographic film,\cite{Rontgen1896OnRays} which required very long exposure and acquisition times. Scintillators became the workhorse detection technique in x-ray imaging around the 1990's with the emergence of digital detectors.\cite{Lecoq2016DevelopmentApplications} ICL, or equivalently free-electron scintillation, has remained a technique of interest in gemology\cite{Remond2000ImportanceMaterials} and semiconductor physics,\cite{Yacobi1986CathodoluminescenceSemiconductors} with applications in cathode-ray tube instruments.\cite{Garlick1950Cathodoluminescence} ICL, scintillation and their applications to nanophotonics are discussed in section~\ref{sec:theory-scint}.

All of the original observations discussed up to this point were performed with bulk media and high-energy electrons. The first occurrence of free-electron radiation in structured media was done in 1953 by Smith and Purcell who observed visible radiation from $\approx$300~keV electrons flying above a metallic diffraction grating.\cite{Smith1953} The effect, now coined as SPR is also sometimes referred to as a form of "diffraction radiation\cite{Potylitsyn2010}". SPR has found direct applications in microwave electronics\cite{Tsimring2006ElectronElectronics} and is considered as a promising platform for non-invasive particle beam diagnosis.\cite{Lampel1997CoherentDiagnostic} SPR is discussed in greater depth in section~\ref{sec:theory-sp}.

The core of this Review is to discuss recent developments in nanophotonics which have enabled a plethora of new effects and a new framework to understand free-electron emission. Specifically, we discuss how the interplay of free-electron physics (and more generally high-energy physics in the case of scintillation) has enabled the control and enhancement of the above-mentioned emission effects. 

\subsection{Recent milestones enabled by nanophotonics}

\begin{figure*}
\includegraphics{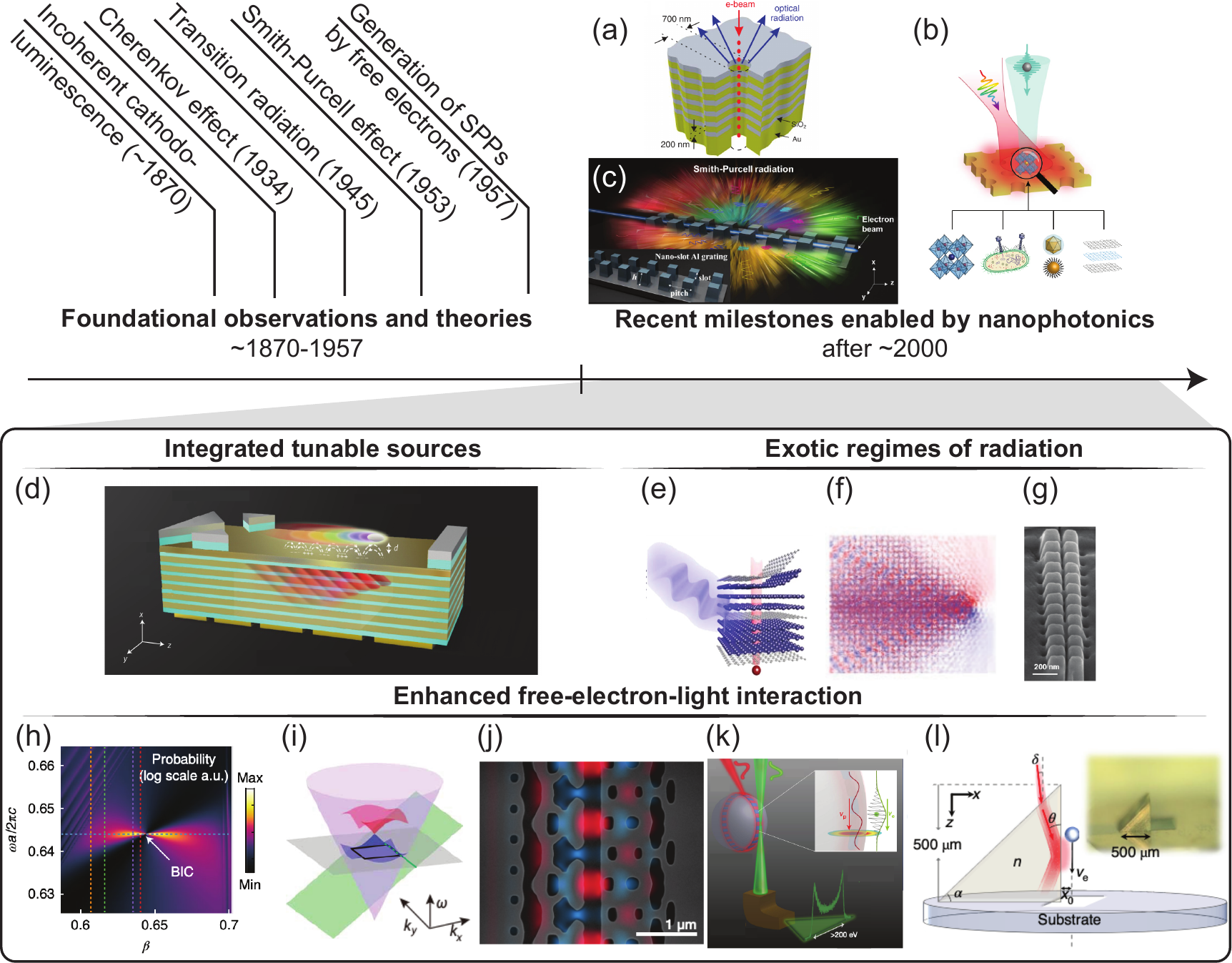}
\caption{\label{fig:timeline} \textbf{Historical timeline: from fundamental discoveries to novel applications enabled by nanophotonics.} The field of free-electron-light interactions was enabled by a few late-19$^\text{th}$ and 20$^\text{th}$-century fundamental discoveries and theoretical proposals such as incoherent cathodoluminescence, CR, TR, Smith-Purcell radiation (SPR), and excitation of surface plasmon polaritons (SPPs) by free electrons. (a-m) Recent advances in nanophotonics have enabled the demonstration of several new effects and technologies relying on those fundamental effects. (a) Light well: tunable electron beam source, reproduced with permission from Ref.~\cite{Adamo2009LightChip}. (b) Ultrafast
multidimensional spectroscopy and microscopy in a transmission electron microscope (TEM), such as perovskite materials, frozen cells, quantum dots and two-dimensional materials, reproduced with permission from Ref.~\cite{Wang2020CoherentCavity}. (c, g) Deep ultraviolet SPR enabled by nano-slot aluminum grating, reproduced with permission from Ref.~\cite{Ye2019Deep-ultravioletRadiation}. (d) Thresholdless integrated Cherenkov emitter in a hyperbolic metamaterial, reproduced with permission from Ref.~\cite{Liu2017IntegratedThreshold}. (e) Tunable x-ray generation from van der Waals heterostructures, reproduced with permission from Ref.~\cite{Shentcis2020TunableMaterials}. (f) Backward Cherenkov emission in PhCs, reproduced with permission from Ref.~\cite{Luo2003CerenkovCrystals}. (h) Bound state in the continuum enhancement of free-electron emission, reproduced with permission from Ref.~\cite{Yang2018MaximalElectrons}. (i) Flatband resonance enhancement of free-electron emission, reproduced with permission from Ref.~\cite{Yang2021ObservationResonances}. (j) Inverse-designed electron accelerator, reproduced with permission from Ref.~\cite{Sapra2020On-chipAccelerator}. (k) Resonant interaction between an electron beam and a photonic resonator, reproduced with permission from Ref.~\cite{Kfir2020ControllingModes}. (l) Phase-matched interaction between a free-electron beam and a light wave, reproduced with permission from Ref.~\cite{Dahan2020ResonantWavefunction}.}
\end{figure*}

A historical timeline of free-electron radiation, from the discovery of its fundamental building blocks to recent effects enabled by nanophotonics, is shown in Fig.~\ref{fig:timeline}. The first wave of discoveries in the field of free-electron radiation happened in the years between 1870-1953. In this period, the fundamental mechanisms were first observed and explained.
What we have seen in the past 10-20 years is a second wave of discoveries, mostly driven by the possibilities and new concepts emerging from the field of nanophotonics. \footnote{This Review focuses on spontaneous emission from free electrons, which is why we do not display major achievements like the first operation of a free-electron laser,\cite{Deacon1977FirstLaser} or work on dielectric laser accelerators\cite{England2014} in the timeline (though we discuss their connection with nanophotonics in section~\ref{sec:future}).}

Around the turn of the 21$^\text{st}$ century, advances in nanofabrication triggered a renewal of interest in understanding light propagation in patterned materials on the scale of optical wavelengths.\cite{Novotny2009PrinciplesNano-optics} In particular, the birth of PhCs enabled an abundance of techniques to control photonic properties in engineered materials.\cite{Joannopoulos2011,Yablonovitch1994PhotonicCrystals} The field of nanophotonics recently found applications in free-electron physics, as a way to control and enhance radiation from free electrons. Alternatively, free electron beams can be used as diagnostic tools to probe photonic properties of nanostructures.\cite{GarciaDeAbajo2010OpticalMicroscopy,Polman2019Electron-beamNanophotonics} 

Perhaps the most obvious, yet much awaited application of nanophotonics in free-electron physics is the miniaturization of free-electron-driven radiation sources.\cite{Ishizuka2001, Roques-Carmes2019TowardsSources, Massuda2017, Liu2017IntegratedThreshold, Adamo2009LightChip,Adamo2012,Adamo2010TuneableSource} Specifically, nanophotonic structures have been shown to allow visible radiation from relatively slow electrons ($\beta<0.2$, which can be generated and accelerated on chip-scale distances), and to eliminate emission threshold in CR.\cite{Liu2017IntegratedThreshold, Vick2018ExtremeSilicon, Zhang2021High-efficiencyBand} This effort has also been bolstered by advances in integrated free-electron sources such as field emitter arrays.\cite{Guerrera2016, Guerrera2016a, Temple1999} 

More generally, nanophotonics offers a convenient platform to control and enhance radiation by engineering the interaction of electron beams with photonic modes.\cite{Yang2018MaximalElectrons, Yang2021ObservationResonances, Remez2017SpectralRadiation, Chuang1984, Coenen2011, Yang2018ManipulatingMetasurfaces} Some regimes of electron emission forbidden in most macroscopic media\cite{Veselago1968ElectrodynamicsPermeabilities} are realizable in some specific nanophotonic structures, such as backward CR in PhC.\cite{Luo2003CerenkovCrystals,Lin2018ControllingRadiation}

One of the most promising advantages of free-electron radiation is its wide tunability and the available wavelength ranges, from microwave to x-ray radiation. This tunability is achieved via structural and electron beam engineering. This is in contrast with wavelength tunability in, e.g. laser sources, which typically requires the sometimes painstaking development of new materials emitting at the wavelength of interest. Specifically, nanophotonic structures pumped by free-electron beams have been shown to emit photons in hard-to-reach regimes, such as UV,\cite{Ye2019Deep-ultravioletRadiation, Vick2018ExtremeSilicon, Watanabe2009Far-ultravioletNitride,Watanabe2011HexagonalApplication,Watanabe2004} soft x-ray,\cite{Shentcis2020TunableMaterials} THz,\cite{Urata1998SuperradiantEmission} and mm-wave.\cite{Goldstein1998} Free electrons provide a versatile platform to access parts of the electromagnetic spectrum where few sources are available, utilizing the radiation control and enhancement techniques mentioned above.

Given the nanometer scale spatial resolution of electron beams in most electron microscopes, free-electron radiation has been considered as a spectroscopic probe to study nanophotonic structures.\cite{Coenen2017CathodoluminescenceLight,GarciaDeAbajo2010OpticalMicroscopy,Polman2019Electron-beamNanophotonics} Free electrons can also interact with nanophotonic structures over extended interaction lengths, thereby achieving stronger coupling strengths, enabling regimes of multi-photons stimulated emission and absorption by a single electron,~\cite{Dahan2020ResonantWavefunction} and electron acceleration in integrated dielectric laser accelerators.~\cite{Sapra2020On-chipAccelerator}

Some of the above-mentioned applications have been enabled by recent advances in ultrafast electron microscopy, where electron-beam and optical excitation of the sample can be modulated in time down to attosecond pulses,\cite{Morimoto2018DiffractionTrains,Priebe2017AttosecondMicroscopy,Vanacore2018AttosecondFields, Kozak2018PonderomotiveTrains} thereby unveiling quantum properties of electron-light interactions.\cite{Feist2015QuantumMicroscope,DiGiulio2019ProbingElectrons, Lim2021ControlWavepackets} In particular, in photon-induced near-field electron microscopy\cite{Barwick2009Photon-inducedMicroscopy,Park2010Photon-inducedExperimental,GarciaDeAbajo2010MultiphotonFields} (PINEM), one can probe near-field non-equilibrium properties of physical systems with unprecedented time and spatial resolution. The recent introduction of nanophotonics in PINEM has enabled the implementation of cavity quantum electrodynamics\cite{Wang2020CoherentCavity,Kfir2020ControllingModes} with free electrons, the generation of electron vortex beams,\cite{Vanacore2019UltrafastFields} the coherent control of electron beam statistics,\cite{Dahan2021ImprintingElectrons} electron beam modulation with silicon photonics,~\cite{Henke2021IntegratedModulation} coincidence electron-photon detection,~\cite{Feist2022Cavity-mediatedPairs, Varkentina2022CathodoluminescencePathways} and strong coupling in the single-photon--single-electron regime.\cite{Adiv2022ObservationRadiation}

\begin{table}
\caption{\label{tab:acronyms} Table of acronyms used in this paper.}
\begin{ruledtabular}
\begin{tabular}{cc}
Acronym& Meaning\\
\hline
CR & Cherenkov effect\\
SPR & Smith-Purcell radiation\\
TR & transition radiation\\
CL & cathodoluminescence\\
ICL & incoherent cathodoluminescence\\
PhC & photonic crystal\\
BIC & bound state in the continuum\\
SPP & surface plasmon polariton\\
BZ & Brillouin zone\\
UV & ultraviolet\\
NIR & near-infrared\\
PINEM & photon-induced near-field electron microscopy\\
EELS & electron energy-loss spectroscopy 
\end{tabular}
\end{ruledtabular}
\end{table}

\section{Typology of free-electron radiation}
\label{sec:theory}


In this section, we describe the basic organizing principles of this Review, which help to sort out the different effects under a general formalism that highlights the role and prospects of nanophotonics in this field.

\begin{figure*}
\includegraphics[scale=0.8]{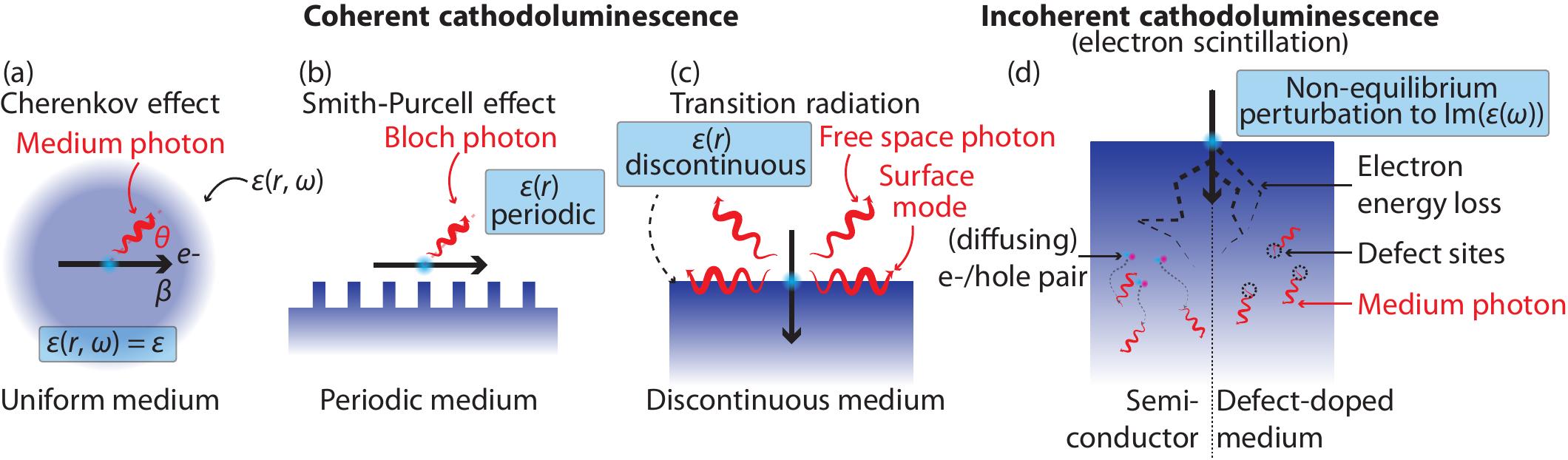}
\caption{\label{fig:nomenclature} \textbf{Examples of spontaneous emission from free electrons.} Free electrons can emit photons spontaneously in several manners. (a) CR: a free electron propagating in a bulk medium at a speed larger than the speed of light in the medium ($c/\text{Re}(\sqrt{\epsilon})$). (b) SPR: in a periodic medium, free electrons can emit a Bloch photon. (c) TR: when going through a discontinuous interface between two media (such that $\epsilon(r)$ is discontinuous), free electrons can emit a surface plasmon at the interface. (d) Incoherent cathodoluminescence (ICL), also known as electron scintillation, occurs in materials with radiating energy levels, such as in semiconductors or in defect-doped media. A free electron impinging on the material can convert part of its energy into a cascade of secondary electrons and excite radiating sites, such as diffusing electron-hole pairs or excited states in implanted defects. In such materials, optical emission can be modeled as a small (perturbative) non-equilibrium loss function $\text{Im}(\epsilon(\omega))$ at the emission wavelength. This review focuses on types of spontaneous emission that have been enhanced or controlled using nanophotonics. Other types of spontaneous emission from free electrons, not covered in this Review, include bremsstrahlung, betatron radiation, and undulator radiation.}
\end{figure*}

\subsection{Coherent vs. incoherent cathodoluminescence}
\label{sec:typology}

All of the effects in the grazing-angle interaction and impact interaction zones have a common physical origin: they result from the coherent interaction between the photonic eigenmodes of the structure and a current source $\mathbf{J}(\mathbf{r},t)$ (describing the particle trajectory). The current distribution can be equivalently described in time and frequency domain. In the following, we consider a trajectory of the form $\mathbf{r}(t) = \mathbf{v}t$ (with the initial position at $t=0$ taken to zero without loss of generality):
\begin{align}
    \label{eq:detcurr}
    \mathbf{J}(\mathbf{r}, t) & = q \mathbf{v} \delta(\mathbf{r}-\mathbf{v}t) \\ 
    \label{eq:detcurr_freq}
    \mathbf{J}(\mathbf{r}, \omega) & = q \hat{r}_\parallel \delta(\mathbf{r}_{\perp}) e^{i\omega r_\parallel/v},
\end{align}
where $q$ is the electron charge. The particle propagates along the linear trajectory defined by the velocity vector $\mathbf{v}=v\hat{r}_{\parallel}$. The unit vector parallel to $\mathbf{v}$ is $\hat{r}_\parallel$ and the orthogonal space is denoted as $\{ \hat{r}_{j,\perp}\}_{j=1,\ldots N-1}$ (where $N$ is the dimensionality of space). Given the mismatch in energies between the incoming electron ($>1$~keV in most settings) and the emitted photon energy (few eV's in the visible to near-infrared (NIR)), one can often safely neglect the quantum emission recoil, and therefore consider the trajectory to be unaffected by radiation processes. 

This current source can be plugged into Maxwell's equations in free space, resulting in an evanescent near field,~\cite{JohnDavidJackson1999ClassicalElectrodynamics} which can be scattered by the structure\cite{GarciaDeAbajo2010OpticalMicroscopy,Polman2019Electron-beamNanophotonics}. This observation also led to the application of fundamental bounds on free-electron radiation and energy loss (derived in section~\ref{sec:theory-bounds} and Ref.\cite{Yang2018MaximalElectrons}). Said differently, the current source (representing the free electron) is performing work on the system and some of which results in radiation. 

The current distribution in Eq.~(\ref{eq:detcurr}) can be modeled in Maxwell's equations in several ways, depending on the type of numerical solver. For instance, in finite-difference time-domain solvers, it can be modeled as an array of dipoles "turned on" sequentially at a speed corresponding to the flight of the electron.\cite{Roques-Carmes2019TowardsSources, Massuda2017} Alternatively, it can be directly injected as a line current in frequency domain (with Bloch periodic boundary conditions to model periodic systems).~\cite{Yang2018MaximalElectrons, Szczepkowicz2020Frequency-domainGratings} 

Coherent CL effects are often considered in opposition to ICL, which originates from stochastic energy losses in the material. Because of the stochastic nature of the process, emitted photons lose their coherence with respect to the incoming electron. Instead of the deterministic current distribution from Eq.~(\ref{eq:detcurr}), one models ICL as radiation from a stochastic current distribution, whose current-current correlations are prescribed as:
\begin{equation}
\label{eq:randcurr}
  \langle J_j^{-}(\mathbf{r}_1,\omega)J_k^{+}(\mathbf{r}_2,\omega) \rangle 
  \equiv 2\pi T S_{jk}(\mathbf{r_1},\mathbf{r}_2,\omega),
\end{equation}
with $S_{jk}(\mathbf{r_1},\mathbf{r}_2,\omega)=\sum_{\alpha,\beta} J_j^{\alpha\beta}(\mathbf{r}_1)J_k^{\beta\alpha}(\mathbf{r}_2) \times f_{\alpha}(1-f_{\beta})\delta(\omega-\omega_{\alpha\beta})$, and $T$ is a normalization time. In this spectral function, $f_{\alpha}$ is the occupation factor of microscopic state $\alpha$ with energy $E_{\alpha}$, $J^{\alpha\beta}$ represents the matrix element of the current density operator $\mathbf{J} \equiv \frac{e}{m}\psi^{\dagger} (-i\hbar\nabla) \psi$, and $\omega_{\alpha\beta} = [E_{\alpha}-E_{\beta}]/\hbar$. The current density matrix and the occupation functions can depend on position, as they depend on the electron energy loss density.

Direct calculation of radiation from the stochastic current source described in Eq.~(\ref{eq:randcurr}) is a computationally expensive problem, as is known in the context of thermal emission.\cite{Chan2006DirectSlabs} Such calculations would indeed require the sampling of a three-dimensional current distribution, whose correlations partially depend on the microscopics and electron energy loss dynamics. Therefore, it is strongly beneficial to resort to more efficient numerical methods, leveraging electromagnetic reciprocity, to make such calculations tractable in three-dimensions.\cite{Roques-Carmes2022ANanophotonics}

The distinction between coherent and incoherent CL is also linked to the final quantum state in which the sample is left after radiation.\cite{GarciaDeAbajo2010OpticalMicroscopy} In the following, we show how CR, TR, SPR, and other coherent CL effects arise from the coherent interaction of the current distribution from Eq.~(\ref{eq:detcurr}) with photonic eigenmodes. We also show how the stochastic current distribution from Eq.~(\ref{eq:randcurr}) can radiate ICL in arbitrary nanophotonic environments, and computational techniques to calculate it efficiently.

\subsection{A unifying picture of coherent cathodoluminescence in arbitrary nanophotonic environments}

Considering the current source in Eq.~(\ref{eq:detcurr}) as a source in Maxwell's equations, one can calculate radiation from a moving free electron in arbitrary nanophotonic media. We expand the Green dyadic tensor -- relating currents to fields linearly as $\mathbf{E}(\mathbf{r},\omega) = i\omega \mu_0 \int d\mathbf{r}~\mathbf{G}(\mathbf{r}, \mathbf{r}',\omega) \mathbf{J}(\mathbf{r}', \omega)$ -- over its set of  eigenmodes\cite{Novotny2009PrinciplesNano-optics} $\{ \mathbf{F}_m (\mathbf{r}, \omega)\}_m$ :
\begin{equation}
    \label{eq:green}
    \mathbf{G}(\mathbf{r}, \mathbf{r}', \omega) = c^2 \sum_m \left( \mathbf{F}_m (\mathbf{r}, \omega) \otimes \mathbf{F}_m^* (\mathbf{r}', \omega) \right) g(\omega, \omega_m),
\end{equation}
where $g(\omega, \omega_m) = (\omega^2 -\omega_m^2 -2i\omega_m\gamma_m)^{-1}$. The eigenmodes $\mathbf{F}_m$ are normalized such that:$\int_V \frac{1}{2\omega_m} \frac{d}{d\omega} \left( \epsilon(\omega) \omega^2 \right) | \mathbf{F}_m (\mathbf{r}, \omega)|^2 = 1$. The main assumption in deriving this equation are small losses and weak dispersion, which are valid in most cases we consider in this work, and can be relaxed further by using the dyadic Green tensor directly\cite{Rivera2020LightmatterQuasiparticles} (without referring to a mode expansion).

The total energy radiated by a dipole can be calculated as an integral in frequency domain,\cite{Kremers2009TheorySpectrum} and we get the following expression:
\begin{widetext}
\begin{equation}
    U = \overbrace{\frac{\mu_0 q^2 c^2}{\pi}}^{\text{prefactor}}  \int dr_\parallel dr'_\parallel \text{Im} \int d\omega ~ \omega \sum_m \underbrace{\big( \hat{r}_\parallel \cdot \mathbf{F}_m(r_\parallel, \mathbf{r}_\perp, \omega) \big) \big( \hat{r}_\parallel \cdot \mathbf{F}^*_m(r'_\parallel, \mathbf{r}_\perp, \omega) \big)}_{\text{mode-electron overlap}} \overbrace{e^{i\omega(r'_\parallel-r_\parallel)/v}}^{\text{phase-matching}} \underbrace{g(\omega, \omega_m)}_{\text{spectral dep.}}.
    \label{eq:genform}
\end{equation}
\end{widetext}
Eq.~(\ref{eq:genform}) already highlights the main physical parameters relevant to understanding all effects in coherent CL, and we therefore refer to it as the "master equation". The prefactor is always proportional to $q^2$, which underlines possible radiation enhancements by considering highly-charged particles (such as heavy ions\cite{Roques-Carmes2018NonperturbativeEffect}). More generally, the $q^2$ dependence offers a mechanism to distinguish between particles with elementary charge $q$ (such as electrons, protons, etc.) and nuclei with charge $Zq$ (where $Z$ is the atomic number). For instance, even CR from fully ionized helium is four times stronger than that of hydrogen isotopes and elementary charges.~\cite{Ginzburg1996RadiationPhenomena} In heavier materials, the larger discrepancy can be used to "count" the energy of each incoming particle.~\cite{Kleinknecht1982ParticleDetectors} 

The rest of the expression is summed over all photonic modes indexed by $m$ which may, in principle, contribute to radiation. The photonic mode-electron overlap highlights the importance of the spatial overlap between the photonic mode and the electron beam (since $\mathbf{F}_m$ is evaluated at the fixed location $\mathbf{r}_\perp$ perpendicular to the axis of propagation of the electron). Radiation might also be enhanced by considering extended interactions over lengths such that $\mathbf{F}_m(r_\parallel, \mathbf{r}_\perp, \omega)$ remains large over the polarization parallel to the electron trajectory. 

Further physical considerations can be made to evaluate the mode-electron overlap. The mode profile $\mathbf{F}_m(r_\parallel, \mathbf{r}_\perp, \omega)$ has a given polarization distribution, but only the polarization along the beam propagation contributes to emission. The spatial dependence of the field profile results in evanescent coupling strengths in many scenarios, as will be made evident for excitation of SPPs and electromagnetic bounds on coherent CL (sections~\ref{sec:theory-spp}~and~\ref{sec:theory-bounds}).

The spectral dependence of the emitted energy is encoded in $g(\omega,\omega_m)$, such that $\text{Im} (g) \approx \pi \delta(\omega^2 - \omega_m^2) \approx \frac{\pi}{2\omega_m}\delta(\omega-\omega_m)$ in the limit of small losses (where $\text{Im}$ denotes the imaginary part). Therefore, radiation in eigenfrequencies of the photonic structure are strongly enhanced. 

The term labeled "phase-matching" in the master equation Eq.~(\ref{eq:genform}) only contributes partially to the phase-matching condition described in Eq.~(\ref{eq:phase-matching}) and is complemented by a phasor of the form $e^{ik_\parallel r_\parallel}$ from the mode-electron overlap term, which for instance appears in systems exhibiting translational invariance (see below). Their combination yields the phase-matching relation from Eq.~(\ref{eq:phase-matching}). To further outline the physical importance of phase-matching, we consider a specific mode with longitudinal wavevector $k_\parallel$, such that $\mathbf{F}_m(r_\parallel, \mathbf{r}_\perp, \omega)~=~e^{ik_\parallel r_\parallel} \mathbf{f}_m(\mathbf{r}_\perp, \omega)$. This yields a version of the master equation where the phase-matching condition is evident:
\begin{equation}
    U \propto \sum_m \underbrace{\big| \hat{r}_\parallel \cdot \mathbf{f}_m(\mathbf{r}_\perp, \omega_m) \big|^2}_{\text{mode-electron overlap}} \overbrace{\Lambda^2\text{sinc}^2\left(\left(\frac{\omega_m}{v}-k_\parallel\right)\Lambda\right)}^{\text{phase-matching}},
    \label{eq:genform_simplified}    
\end{equation}
where $\text{sinc}(x) = \sin(x)/x$ is the sinc function, and $\Lambda$ is the length of interaction. The phase-matching term becomes a Dirac delta function for large interaction lengths $\Lambda$, thereby highlighting the critical importance of phase-matching in physical settings with extended interactions. Compared to Eq.~(\ref{eq:genform}), this equation also highlights the spectral dependence of the radiation, which is determined by the photonic eigenfrequencies $\omega_m (\mathbf{k})$. This compact form also hints at a geometric method to calculate the emitted power by (1) considering the intersection of the band structure dispersion $\omega_m (\mathbf{k})$ with the electron plane $\mathbf{k}\cdot \mathbf{v}=k_\parallel v$; (2) weighting each intersection by the mode-electron overlap term in Eq.~(\ref{eq:genform_simplified}). We described in section~\ref{intro-pm} how such a method can be used to predict radiation in various radiation events occuring in the grazing-angle interaction zone in the introduction, for CR in bulk media and PhCs, SPR, and excitation of SPPs. 

To gain further physical insight with this formalism, we must consider specific photonic environments, which will be described by various eigenmode distributions, and are discussed in the next sections.

\subsubsection{Cherenkov radiation}
\label{sec:theory-cherenkov}

CR in its simplest embodiment occurs in a homogeneous dielectric medium (of index $n$) and consists in the spontaneous emission of plane waves by a charged particle (see Fig.~\ref{fig:nomenclature}(a)). Its description, in the language of the previous section and the master equation, can be understood in terms of plane wave eigenmodes of a medium of index $n$. The dispersion relation is $\omega_k = c|\mathbf{k}|/n$. Plugging this expression into the master equation Eq.~(\ref{eq:genform}), we can get the famous \emph{Frank-Tamm formula}~\cite{Frank1937CoherentMatter} for the spectral density per unit propagation length, shown in the Appendix~\ref{sec:appendix-cherenkov}. Emission is only allowed for superluminal electrons, which is equivalent to the phase-matching condition from Eq.~(\ref{eq:phase-matching}) in a bulk medium. 

Quantum corrections to this formalism can be introduced, taking into account recoil,\cite{Ginzburg1940QuantumMedium, Cox1944MomentumRadiation} non-perturbative effects with heavy ions,\cite{Roques-Carmes2018NonperturbativeEffect} the particle's spin and orbital angular momentum,\cite{Kaminer2016QuantumMomentum} reduced dimensionality,\cite{Adiv2022ObservationRadiation} or emission from hot electrons in two-dimensional materials.\cite{Kaminer2016EfficientGraphene}

\subsubsection{Smith-Purcell radiation}
\label{sec:theory-sp}

SPR is a natural extension of CR to periodic media as the spontaneous emission of Bloch photons\cite{Rivera2020LightmatterQuasiparticles} (see Fig.~\ref{fig:nomenclature}(b)), which we use as eigenmodes in the expansion from Eq.~(\ref{eq:green}). We first consider the case of a one-dimensional periodic structure (period $L$) along the direction of electron propagation. Eigenmodes can be described by the band index $m$ and reciprocal lattice vector $G$ in the first BZ, such that $\mathbf{F}_{m,k_\parallel} = \ \sum_G \mathbf{c}_{m,k_\parallel}^G (\mathbf{r}_\perp) e^{-ir_\parallel (G+k_\parallel)}/\sqrt{L}$, where $\mathbf{c}_{m,k_\parallel}^G$ are coefficients of the mode's Fourier expansion. 

We then get the following expression for the energy spectral density:
\begin{equation}
    \label{eq:sp1d}
    \frac{1}{L}\frac{dU}{d\omega} = \frac{\mu_0 q^2 c^2}{2} \sum_{m}\sum_G \left| \hat{r}_\parallel \cdot \mathbf{c}_{m,k_\parallel}^G (\mathbf{r}_\perp) \right|^2 \delta(\omega - \omega_{m,k_\parallel}).
\end{equation}
The $\delta$ function in Eq.~(\ref{eq:sp1d}) gives us a geometric way of predicting which Bloch modes are excited by the electron beam, by considering the intersection of the band structure $\omega_{m,k_\parallel}$ with $\omega = k_\parallel v$. This method generalizes well to higher dimensions where the electron "line" becomes a plane in the band structure representation (see Fig.~\ref{fig:overview}(b,d) for some example band structures). The phase-matching condition also sets the dispersion relation, known as the Smith-Purcell relation: 
\begin{equation}
    \label{eq:spdisp}
    \lambda = \frac{L}{m}\left(\frac{1}{\beta} - \cos \theta\right),
\end{equation}
where $\lambda$ is the photon wavelength, and $m$ the Bloch mode index. The emission angle $\theta$ is measured, as in the case of CR, with respect to the electron propagation direction. 

\begin{figure}
\includegraphics[scale = 0.35]{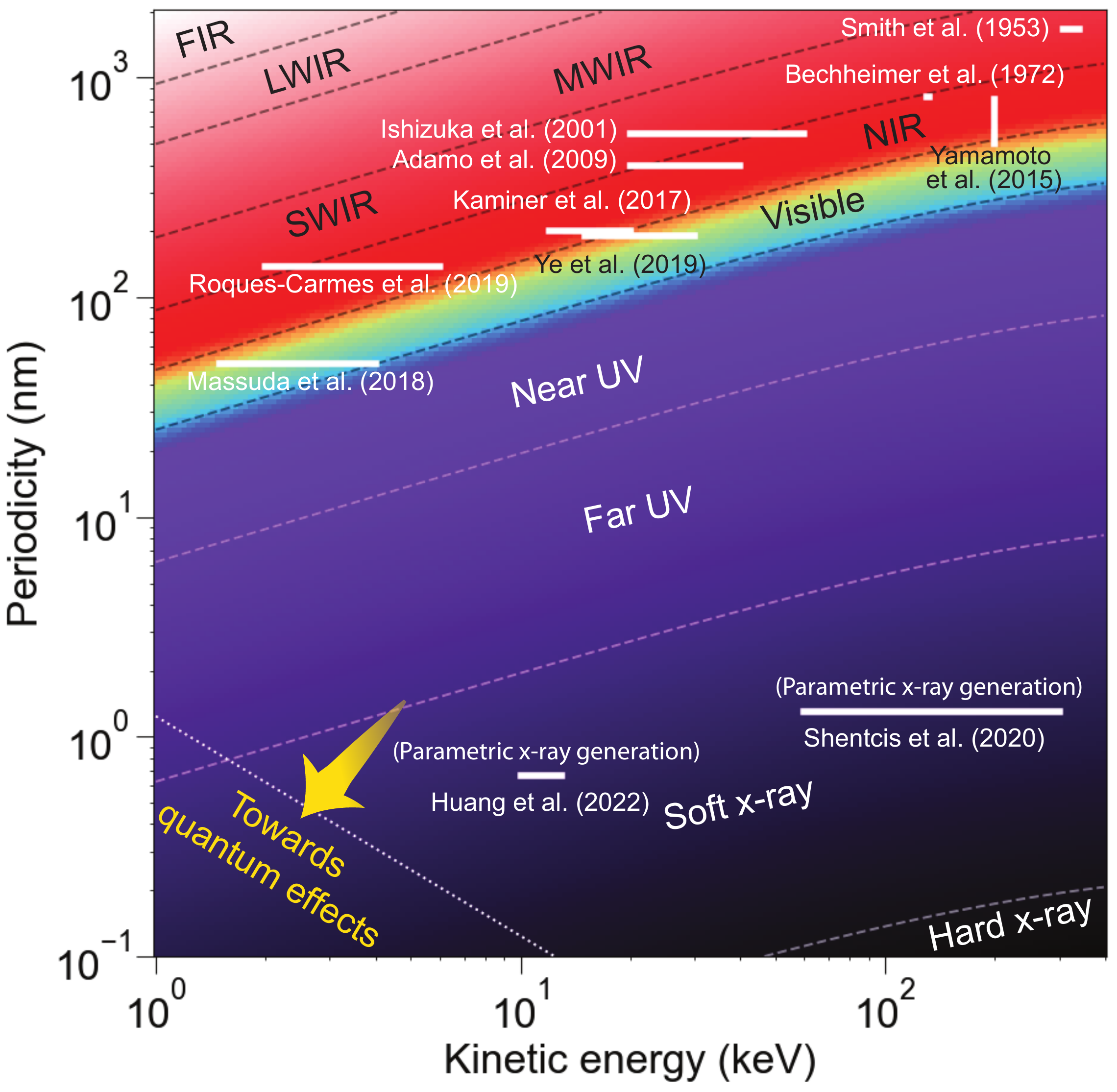}
\caption{\label{fig:spoverview} \textbf{Overview of progress in Smith-Purcell emitters.} Each straight thick white line corresponds to a reported experimental observation, starting from the original observation from Smith and Purcell\cite{Smith1953} (top right corner). The overview highlights recent progress enabled by nanophotonics towards high-frequency radiation (visible and ultraviolet) with short-pitch gratings and/or low-energy electrons (<10~keV). The background color corresponds to the emitted wavelength "color" at normal incidence $\theta = 90^\circ$. References appearing in the figure are Refs.\cite{Smith1953, Bachheimer1972ExperimentalEffect, Yamamoto2015, Ishizuka2001, Adamo2009LightChip, Kaminer2017SpectrallyDisorder, Ye2019Deep-ultravioletRadiation, Roques-Carmes2019TowardsSources, Massuda2017, Shentcis2020TunableMaterials, Huang2022QuantumCrystals}. In some references,\cite{Ye2019Deep-ultravioletRadiation, Bachheimer1972ExperimentalEffect} the main reported result comes from a higher-order Smith-Purcell peak, which is why the work appears in a longer wavelength domain (e.g., Ref.\cite{Ye2019Deep-ultravioletRadiation} reports radiation in the deep ultraviolet with up to third-order SPR). In some other references,\cite{Smith1953} radiation is measured at an angle closer to $\theta = 0^\circ$, therefore reporting shorter wavelengths than what is displayed here. In Ref.\cite{Shentcis2020TunableMaterials, Huang2022QuantumCrystals}, free electrons emit soft x-ray photons via parametric x-ray radiation or coherent bremsstrahlung. UV: ultraviolet; NIR: near infrared; SWIR: short wave infrared; MWIR: mid wave infrared; LWIR: long wave infrared; FIR: far infrared. "Quantum effects" denote the region where recoil effects become observable, delimited by $\lambda (\theta = 90^\circ) = \lambda_e$, where $\lambda (\theta = 90^\circ)$ is the emission wavelength at normal incidence, and $\lambda_e$ the de Broglie wavelength of the electron.}
\end{figure}

This simple relation enables us to make quick predictions of the radiation spectrum (in the absence of resonant enhancement), knowing the structure periodicity along the electron trajectory. Several observations of SPR from various periodicities and electron energies are shown in Fig.~\ref{fig:spoverview}, on a background corresponding to the wavelength predicted by Eq.~(\ref{eq:spdisp}) with $\theta=\pi/2$ and $m=1$. The original observation from Smith and Purcell was performed with $\beta \approx 0.78-0.8$. The relatively large grating pitch they used should have resulted in radiation in the short-wave infrared at normal emission direction, but they measured visible radiation at a shallower angle $\theta \approx 20^\circ$. 

Early work on SPR was essentially focused on metallic structures.\cite{Salisbury1970GenerationElectrons} More recent work used electrons in similar energy ranges, pushing radiation into the near-UV regime with higher diffraction orders.\cite{Yamamoto2015} With the goal of integrating Smith-Purcell emitters into optoelectronic devices, recent effort has been targeted at reducing the electron kinetic energy and periodicity, to retain emission in the NIR to UV regimes\footnote{We note that SPR was also observed with relativistic electrons.\cite{Doucas1992FirstElectrons}}. This effort has been enabled by recent progress in nanofabrication, namely the capability of fabrication sub-100~nm periodic structures with electron-beam lithography.\cite{Massuda2017} Some works even reported emission from electron beams generated on-chip by field emitters\cite{Ishizuka2001} and/or with integrated all-silicon structures.\cite{Roques-Carmes2019TowardsSources} The prospect of generating radiation deeper in the UV remains an exciting perspective, with recent work demonstrating SPR in the UV with $\beta\approx0.24-0.33$ and leveraging higher-order diffraction modes.\cite{Ye2019Deep-ultravioletRadiation} Beyond the technological promise of integrated and tunable UV emitters, short-wavelength SPR also exhibits quantum recoil effects in low-energy and short-period settings.\cite{Tsesses2017LightNanostructures} While demonstrations of short-wavelength SPR have been limited to the near UV, other free-electron radiation effects, such as coherent bremsstrahlung and parametric x-ray radiation, being similar in some respects to SPR, are amenable to generation of x-ray radiation. In these effects, the electron flies in a "structured" medium consisting of a crystalline lattice \emph{of atoms}, such that the period is on the \aa ngstr\"{o}m scale. Those techniques offer an interesting platform to generate x-ray photons with moderately relativistic electrons.\cite{Baryshevsky2006CoherentCrystal, Baryshevsky1983ParametricUtilization,Uberall1956High-energyCrystals, Korobochko1965OnBremsstrahlung, Wong2021ProspectsNanomaterials, Tan2021Space-TimeRadiation} We highlight recent work\cite{Shentcis2020TunableMaterials, Balanov2021TemporalInteraction} to reveal the similarities and differences between SPR and the previously-mentioned x-ray emission techniques.

Many theoretical studies focused on SPR and CR in 2D, where the structures are assumed to be invariant along the third dimension and the electron beam is assumed to be a "sheet" beam. In higher dimensions, the point-like nature of the electron must be considered, and the photonic band structure gives greater flexibility in tuning the coupling between the electron and photonic modes. Considering the case of a three-dimensional periodic PhC, we get the following expression, as previously reported in Ref.\cite{Kremers2009TheorySpectrum}:
\begin{equation}
    \label{eq:sp3d}
    \frac{dU}{d\omega dl} = \frac{q^2}{8\pi^2\epsilon_0} \sum_{m, \mathbf{G}} \int_{\partial S} dk \frac{|\mathbf{c}^\mathbf{G}_{m,\mathbf{k}}(\omega)\cdot \hat{r}_\parallel|^2}{|\nabla_{\mathbf{k}_\perp}\omega_{m,\mathbf{k}}|},
\end{equation}
where $\partial S$ is the contour defined by $\omega = \omega_{m,\mathbf{k}}$. This formula shows that the radiated power is proportional to the Fourier coefficient $|\mathbf{c}^\mathbf{G}_{n,\mathbf{k}}(\omega)\cdot \hat{r}_\parallel|$, describing the coupling of the current density with the electromagnetic field at the electron location. Also, the emitted power is proportional to the inverse of the transverse group velocity. This suggests a path to strongly enhance emission from electrons using engineered bands in PhCs.\cite{Kremers2009TheorySpectrum, Yang2018MaximalElectrons, Yang2021ObservationResonances}

\subsubsection{Transition radiation}
\label{sec:theory-tr}
To describe transition radiation within the same framework, we consider the simplified case of a charge impinging at normal incidence on a perfect conductor (see Fig.~\ref{fig:nomenclature}(c)) and resort to the introduction of an image charge with opposite charge and velocity. The current distribution $\mathbf{J}(\mathbf{r},\omega)$ is modified accordingly and its emission in free space is considered. Doing so, we get the spectral distribution shown in Appendix~\ref{sec:appendix-tr}, first derived by Ginzburg and Tamm.\cite{Ginzburg1940QuantumMedium}

This relation can be extended to interfaces between two media with finite permittivities,\cite{Ginzburg1940QuantumMedium, VLGinzburg1984TransitionTheory} radiation into waveguides,\cite{KABarsukov1960TransitionWaveguide} and metallic thin films.\cite{Garibyan1960TransitionLosses} Resonant transition radiation, the coherent interference of multiple TR emission in a multilayered medium, has also been considered as a promising platform for high-energy particle detectors\cite{Lin2018ControllingRadiation,Lin2021ADetectors} and x-ray emission.\cite{Yamada1999ObservationPeriod} Though considered as an extension of TR at a single interface, we note that resonant TR has a dispersion relation similar to that of SPR.\cite{Pardo2000ClassicalStructures}

\subsubsection{Coherent excitation of surface plasmon polaritons (SPPs) by free electrons}
\label{sec:theory-spp}


SPPs can be excited by free electrons~\cite{Keller1994ElectrodynamicWaves, Bashevoy2006GenerationImpact, Liu2012SurfaceSource, GarciiaDeAbajo2013MultipleElectrons, Gong2014ElectronPolaritons, Lin2017SplashingElectrons, Liu2017IntegratedThreshold, Adiv2022ObservationRadiation} either when impinging on metals or when grazing the interface, with first observations reported in Ref.~\cite{Bashevoy2006GenerationImpact} and Ref.~\cite{Adiv2022ObservationRadiation}, respectively. This phenomenon is observed even in the absence of corrugation at the surface, and can be understood with our formalism, when considering SPP eigenmodes. For concreteness, we consider the case of an interface between two media $\epsilon_1>0$ and $\epsilon_2$, with the electron propagating in $\epsilon_1$. The lower branch of the SPP mode is shown in Fig.~\ref{fig:overview}(c) for a Drude-like metal $\epsilon_2 = 1 - (\omega_p/\omega)^2$ (where $\omega_p$ is the plasma frequency). The energy emitted by the free electron per unit length is derived and shown in Appendix~\ref{sec:appendix-spp}.

As expected from the master equation Eq.~(\ref{eq:genform}), the exponential decay of the field at the interface results in a factor $\exp\left({-2|k_\perp|r_\perp}\right)$, arising from the mode-electron overlap evaluated along the electron trajectory. As with other effects in the grazing-angle interaction zone, one can predict which SPP modes are excited by an electron with velocity reduced $\beta$ by identifying the intersection of the dispersion relation with the electron line\cite{Liu2012SurfaceSource} $\omega(k_\parallel) = v k_\parallel$. Analogues of this effect have been observed in systems supporting Dyakonov surface waves\cite{Hu2022SurfaceRadiation} and hyperbolic dispersion.\cite{Ye2019Deep-ultravioletRadiation}

\subsection{Fundamental bounds for coherent cathodoluminescence}
\label{sec:theory-bounds}


Another perspective that the field of nanophotonics brought to research on free-electron radiation is the idea of setting electromagnetic bounds from first principles. Nanophotonics research led to the formulation of universal bounds on various photonic processes,~\cite{Chao2021PhysicalResponse} such as scattering and absorption,~\cite{Miller2016FundamentalSystems} focusing,~\cite{Shim2020MaximalWaves} Raman scattering,~\cite{Michon2019LimitsScatterers} and near-field optical response.~\cite{Shim2019}  The same approach was applied in Ref.~\cite{Yang2018MaximalElectrons} to propose a universal bound on free-electron radiation and energy loss. Such bounds represent the maximal amount of power that could be scattered or absorbed by an optimal structure excited by free electrons and enclosed in a given volume.

Coherent CL can be interpreted as a scattering problem, and is therefore amenable to recent work on electromagnetic bounds.\cite{Yang2018MaximalElectrons} Intuitively, the scattering problem can be bounded by a convex optimization problem, whose solution is obtained by calculating variational derivatives of the incident fields.\cite{Yang2018MaximalElectrons} (The incident evanescent field is generated in free space by the current distribution\cite{GarciaDeAbajo2010OpticalMicroscopy} $\mathbf{J}(\mathbf{r},\omega)$.) Maximal radiation and energy loss powers can then be derived\cite{Yang2018MaximalElectrons} for an arbitrary "scatterer" (corresponding to the sample inducing coherent CL) with susceptibility $\chi(\mathbf{r}, \omega) = \epsilon(\mathbf{r},\omega) -1$ and volume $V$:
\begin{equation}
    \label{eq:bound}
    P_\tau \leq \frac{q^2 \xi_\tau}{8\epsilon_0\omega\pi^2} \int_V d\mathbf{r} \underbrace{\frac{|\chi|^2}{\text{Im}\chi}}_{\text{material}} \overbrace{\left( k_\perp^4 K_0^2(k_\perp \rho) + k_\perp^2 k_\parallel^2 K_0^2(k_\perp \rho) \right)}^{\text{geometry}},
\end{equation}
where $\tau \in \{ \text{rad}, \text{loss} \}$, $\xi_\text{loss} = 1$, $\xi_\text{rad} = \eta (1 - \eta) \leq 1/4$ (with $\eta$ the ratio of radiative to total energy loss), $k_\parallel = \omega/v$, and $k_\perp~=~\sqrt{k_\parallel^2-k^2}=k/\beta\gamma$ (with $\gamma$ the Lorentz factor). $K_n$ is the $n$-th order modified Bessel function of the second kind.

The power bounds from Eq.~(\ref{eq:bound}) apply to the non-retarded or retarded regimes, and only assume the absence of gain in the optical medium. This expression also highlights the main physical parameters of interest to maximize radiation from free electrons: the material factor $|\chi|^2/\text{Im}\chi$ which reflects the influence of material choice, depending on the wavelength of interest; the electron velocity $\beta$ and Lorentz factor $\gamma$ appearing in the impact parameter $k_\perp\rho$ in the integral. Approximations of this bound\cite{Yang2018MaximalElectrons} explicit the role of the minimum distance between the electron trajectory and the scatterer $d$, such that the relevant length scale of interaction is set by $k_\perp d = kd/\beta\gamma$. At large beam-sample distances, the bound decays exponentially $\propto e^{-2k_\perp d}$, which matches the dependence from Eq.~(\ref{eq:spp}). 

The analytical bound also reveals several interesting physical behaviors. Namely, there exists a regime of near-field interaction between the beam and the structure where slow electrons are favored (i.e., they radiate more efficiently). Recent works also highlighted the possibility of strong interactions between slow electrons and plasmonic near fields.~\cite{Talebi2019Near-Field-MediatedInteractions, Talebi2020StrongPrinciples, Liebtrau2021SpontaneousFields} This supports recent interest in developing on-chip sources of free-electron radiation\cite{Ye2019Deep-ultravioletRadiation,Massuda2017,Roques-Carmes2019TowardsSources}, where electron beams can be precisely aligned to nanophotonic structures (e.g. gratings for SPR) to control the beam-sample coupling. 

Another interesting feature is the apparent divergence of the bound in the limit of small losses, which suggests mechanisms to strongly enhance free-electron radiation with high-$Q$ resonances, a path which we discuss in section~\ref{sec:enhance-coherent}.

\subsection{Incoherent cathodoluminescence (electron scintillation)}
\label{sec:theory-scint}

All of the previous types of radiation were forms of coherent CL, which naturally give themselves into control via shaping of the photonic eigenmodes, as we explained in the previous paragraphs. We now consider the case of ICL, or electron scintillation, and propose ways to control and enhance this form of radiation. The method we propose here is readily transferable to scintillation from other types of high-energy particles, such as x- and $\gamma$-rays. 

ICL is usually observed when a beam of electrons is bombarding a material directly (and is therefore occurring in the impact interaction zone), as depicted in Fig.~\ref{fig:nomenclature}(d). Energy is then lost by the electron beam, which can be transferred to radiative sites (electron-hole pairs in semiconductors, or defect states in a doped medium), which subsequently radiate in a nanophotonic environment described by the eigenmode expansion from Eq.~(\ref{eq:green}). 

We can calculate the emitted energy by the stochastic current distribution in Eq.~(\ref{eq:randcurr}), and for simplicity make the assumption that the current correlations are local, isotropic, and real-valued $S_{ij}(\mathbf{r}_1, \mathbf{r}_2,\omega) = \delta_{ij} \delta (\mathbf{r}_1-\mathbf{r}_2)S(\mathbf{r}_1,\omega)$, a condition which can be straightforwardly relaxed.\cite{Roques-Carmes2022} ICL is, in general, described by the light emission from this non-equilibrium steady-state distribution. This assumption is corroborated by the fact that energy deposition occurs on picosecond time scales, which are effectively instantaneous relative to the excited-state depletion time scales (typically nanoseconds).~\cite{Polman2019Electron-beamNanophotonics, Greffet2018LightLaw} We then get:
\begin{equation}
    \label{eq:scint}
    \frac{d\langle U \rangle}{d\omega} = \frac{\pi T}{\epsilon_0} \sum_m \int d\mathbf{r} \left| \mathbf{F}_m(\mathbf{r},\omega)\right|^2 S (\mathbf{r},\omega) \delta(\omega-\omega_m).
\end{equation}
This expression makes explicit the way in which ICL (where the light emission results from spontaneous emission) can be controlled by photonic shaping via the eigenmodes $\mathbf{F}_m(\mathbf{r},\omega)$. It is also apparent that ICL can be enhanced by optimizing the overlap between a given eigenmode and the current correlation function $S(\mathbf{r},\omega)$ (for example, by optimizing the overlap between the photonic eigenmode and the energy-loss density of the high-energy particles).

Another formulation of ICL can be obtained by using electromagnetic reciprocity and the Green's function directly (instead of its eigenmode expansion), yielding the following expression for the power spectrum per unit solid angle and frequency:~\cite{Roques-Carmes2022ANanophotonics}
\begin{equation}
    \frac{dP^{(i)}}{d\omega d\Omega} = \frac{\pi }{\epsilon_0 \omega} \times S(\omega) \times \left[V^{(i)}_\text{eff}(\omega, \Omega)/\lambda^3\right],
    \label{eq:scint-reciprocal}
\end{equation}
where we also assumed that the current correlations are uniform and isotropic in the material (corresponding for example to uniform energy loss), and $V^{(i)}_\text{eff}(\omega, \Omega) = \int_{V_S} d\mathbf{r}~ |\mathbf{E}^{(i)}(\mathbf{r}, \omega, \Omega)|^2 / |\mathbf{E}^{(i)}_\text{inc}(\omega, \Omega)|^2$. Eq.~(\ref{eq:scint-reciprocal}) states that the ICL spectrum, under this approximation, is a simple product of a microscopic factor, set by the non-equilibrium steady-state distribution function $S(\omega)$,~\cite{Kurman2020Photonic-CrystalDetection, Roques-Carmes2022ANanophotonics} and an effective absorption volume $V_\text{eff}$, which is only determined by the (structured) optical medium surrounding the emitting sites.

Eq.~(\ref{eq:scint-reciprocal}) enables a key simplification thanks to electromagnetic reciprocity, which relates the following two quantities: (1) the emitted ICL from the structure (at a given frequency $\omega$, direction $\Omega$, and polarization $i$) and (2) the intensity of the field induced in the sample ($|\mathbf{E}^{(i)}(\mathbf{r}, \omega, \Omega)|^2$) by sending a plane wave at it (of frequency $\omega$, propagating along direction $\Omega$ (with field $\mathbf{E}^{(i)}_\text{inc}(\omega, \Omega)$) into the structure, and polarization $i$). This expression opens the path to efficient numerical methods for ICL and scintillation in three dimensions and arbitrary nanophotonic environments.\cite{Roques-Carmes2022ANanophotonics} 

\section{Control and enhancement of cathodoluminescence with nanophotonics}
\label{sec:expt}

The previous section suggested several ways in which one can control CL with engineered nanophotonic structures. In this section, we describe several methods to experimentally measure (coherent and incoherent) CL from nanophotonic structures in electron microscopes; we also review experimental demonstrations and nanophotonic techniques to control and enhance CL.

\subsection{Cathodoluminescence in electron microscopes}
\label{sec:expt-details}

It is worth noting that CL in electron microscopes was an already-established technique before the 2000's, given its widespread use in gemology and materials science.\cite{Remond2000ImportanceMaterials} The transfer of this technique to the characterization of nanophotonic structures enabled the observation of electron-light interactions in nanophotonic structures via direct spectroscopic techniques (measuring CL) and indirect electron measurement techniques (EELS).

\begin{figure*}
\includegraphics[scale = 0.7]{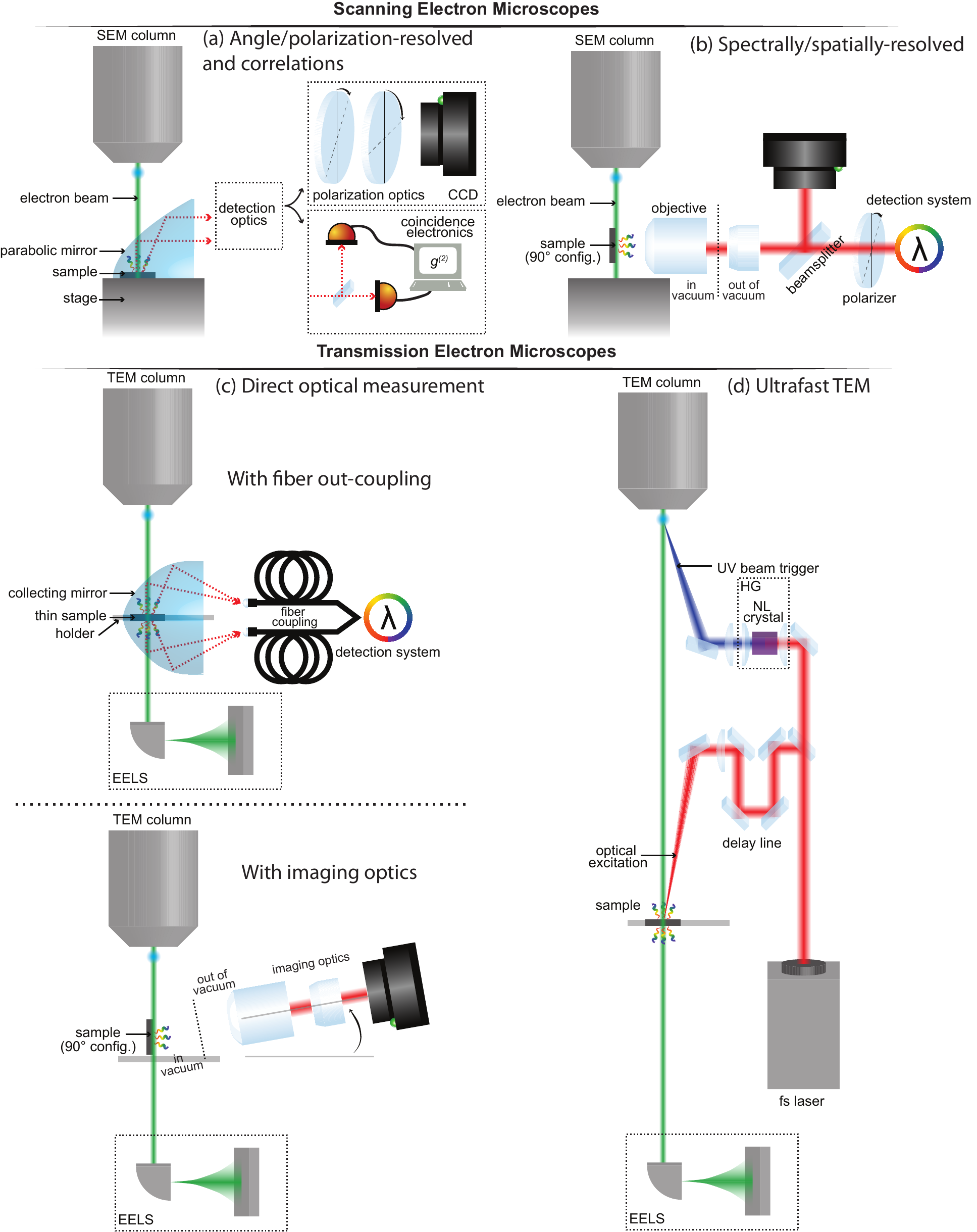}
\caption{\label{fig:setups} \textbf{Probing electron-light interactions in scanning and transmission electron microscopes.} The figure illustrates example systems that represent the wide range of experimental capabilities in modern electron microscopy for measuring free-electron-light interactions in nanophotonics. (a) Angular/polarization-resolved and correlation measurements in a scanning electron microscope.\cite{Coenen2011, Osorio2015Angle-ResolvedPolarimetry, Meuret2015PhotonCathodoluminescence, Sola-Garcia2021Pump-probeMicroscopy, Meuret2019ComplementaryMicroscope} An electron beam impinges on a sample through a hole in a parabolic mirror. Light emitted by the sample upon bombardment is collimated with a parabolic mirror. Light can be (1) analyzed with polarization optics before being detected by a CCD camera or (2) analyzed with a Hanbury Brown-Twiss setup to measure intensity correlations. The electron beam might be triggered by an ultrafast laser or a beam blanker~\cite{Polman2019Electron-beamNanophotonics} (similar to ultrafast TEM shown in (d)). (b) Spectrally-spatially-resolved cathodoluminescence in a scanning electron microscope.\cite{Kaminer2017SpectrallyDisorder, Massuda2017, Roques-Carmes2019TowardsSources, Yang2018MaximalElectrons, Roques-Carmes2022ANanophotonics, Yang2021ObservationResonances, Massuda2017a} Sample is mounted in a 90$^\circ$ configuration. Light emitted from the sample is collected with an objective and imaged on a CCD. Light is also coupled to a spectrometer. Spatial resolution is obtained within the CCD field of view. (c) Direct cathodoluminescence measurement in a transmission electron microscope. Top: with collecting mirrors.\cite{Silver2015CathodoluminescentMicroscope, Stoger-Pollach2017TransitionCathodoluminescence, Stoger-Pollach2019FundamentalsSemiconductors, Stoger-Pollach2021CoherentMicroscope} An electron beam goes through two facets of a collecting mirror and a thin sample. Electron energy loss spectroscopy (EELS) is performed with a magnetic prism after the electron-sample interaction. Light is collected with mirrors focusing light into two optical fibers, which are then analyzed with a detection system. Bottom: with imaging optics.\cite{Remez2017SpectralRadiation, Remez2019ObservingEmission} An objective, tube lens, and CCD rotating assembly is used to measure radiation from a sample in a 90$^\circ$ configuration. (d) Ultrafast transmission electron microscope setup.\cite{Barwick2009Photon-inducedMicroscopy, Piazza2015SimultaneousNear-field, Feist2015QuantumMicroscope, Priebe2017AttosecondMicroscopy, Morimoto2018DiffractionTrains, Vanacore2018AttosecondFields, Vanacore2017, Vanacore2019UltrafastFields, Wang2020CoherentCavity, Kfir2020ControllingModes, Dahan2020ResonantWavefunction, Dahan2021ImprintingElectrons, Henke2021IntegratedModulation} An electron beam is triggered by an ultraviolet pulse generated via a HG (harmonic generation) setup with a NL (nonlinear) crystal pumped by a femtosecond laser. A delay line is used to control the time delay between the optical excitation of the sample and the beam trigger. EELS is also used to measure the electron beam dispersion.}
\end{figure*}

\subsubsection{Cathodoluminescence spectroscopy and polarimetry techniques}
\label{sec:spectroscopy}

Apart from optical-CL instruments $-$ in which an electron beam is generated under moderate vacuum via discharge in a chamber (small enough to be mounted on a standard optical microscope)  $-$ most modern CL instruments are based on a scanning electron microscope (SEM) or a transmission electron microscope (TEM). Fig.~\ref{fig:setups} illustrates representative types of SEM- and TEM-based CL instruments. In all instruments, the light generated by the interaction of a focused electron beam with a sample is out-coupled (with free space collection optics or an optical fiber). Collection optics that have been used for CL measurement and characterization include parabolic mirrors (for angular resolution) and objective lenses (for spatial resolution). Depending on how the light is collected and what additional components are utilized, a wide range of measurements can be performed from obtaining spatially resolved spectral information to angular, polarization, and even time-sensitive detection. Alternatively, placing the nanostructures on the tip of a fiber enables to directly collect the radiation through the fiber, a technique shown to enhance the evanescent field of the free electrons interacting with a nanostructure.~\cite{So2014FiberRange} It has also been demonstrated that the evanescent field of free electrons can by amplified as electrons fly over a plasmonic surface.~\cite{So2015} Having to pass through a window to exit a vacuum chamber can limit the detectable wavelength range. An extensive review of current CL measurement techniques was recently published.\cite{Coenen2017CathodoluminescenceLight} 

Relevant to this Review, we highlight selected papers detailing different experimental methods useful for nanophotonic applications of CL. The use of tightly focused electron beams allows for collecting spectral information directly from nanoscale samples with spatial resolution limited by the size of the focused electron beam. CL found applications in plasmonics,\cite{Bashevoy2006GenerationImpact,Bashevoy2007HyperspectralResolution,Vesseur2007DirectSpectroscopy,Hofmann2007PlasmonicCathodoluminescence, Denisyuk2010TransmittingFeed,Schefold2019SpatialMicroscopy,  Han2018RevealSubnanoscale, Bauman2017PlasmonicImaging, Liu2020MappingMicroscope, Ron2020CathodoluminescenceNetworks, Myroshnychenko2012PlasmonStudy, Krehl2018SpectralResolution} photonics,\cite{Brenny2016Near-InfraredWaveguides, Talebi2015ExcitationResonances}  semiconductors,\cite{Yan2008StructureHeterostructures, Prabaswara2017SpatiallyCathodoluminescence, Vu2022Exciton-dielectricCathodoluminescence} electron-beam lithography,\cite{Edwards2021High-performanceSpectroscopy} and tomographic reconstruction.\cite{Atre2015NanoscaleSpectroscopy} Additionally, the combination of spatial and spectral resolution can be combined to measure the dispersion of CL effects such as SPR.\cite{Kaminer2017SpectrallyDisorder,Massuda2017,Roques-Carmes2019TowardsSources,Yang2018MaximalElectrons} Methods were also proposed to disentangle several types of emission from farfield measurements.\cite{Fiedler2022DisentanglingRadiation} Inspired by the early days of radio, it has been shown that a nanoscopic dipole Herzian antenna acts as an efficient emitter of visible light when an electron beam is injected in the dipole gap.~\cite{Denisyuk2010TransmittingFeed}

Thermal measurements can also be performed with CL, such as nanoscale thermometry and thermal transport measurements both in low current conditions (where sample heating is avoided) and higher current conditions\cite{Mauser2021EmployingNanowires} (where electron beam induced heating is present). Other electron-beam induced effects such as phase transitions in gallium nanoparticles at picojoule excitation energies have also been observed. Transformations between coexisting structural phases are accompanied by continuous changes in the nanoparticle film’s reflectivity~\cite{Pochon2004PhaseExcitation} and luminescence,~\cite{Denisyuk2008LuminescenceState}  which may be used for modulating light and optical data storage. Promising new techniques for resolving and hyperspectrally mapping picometric movements by detecting secondary electron emission from the edge of the nanostructure in an electron microscope~\cite{Liu2021VisualizationNanostructures} might also be realized via cathodoluminescence microscopy in the future.

Next, phase-sensitive imaging measurements were performed utilizing both angle-resolved CL and hyperspectral angle-resolved cathodoluminescence to characterize the far-field phase signal from scattering off of plasmonic nanostructures allowing for the reconstruction of the angle-dependent phase distributions\cite{Schilder2020Phase-ResolvedHolography} and the coupling between nanoparticles and SPPs.\cite{Sannomiya2020CathodoluminescencePolaritons} Similar techniques were used in the Fourier domain to determine the emission polarization properties of sub-wavelength structures like optical nanoantennas.\cite{Coenen2012Polarization-sensitiveMicroscopy} CL measurements have also been used to directly image plasmonic modes in annular nanoresonators, ultrathin plasmonic strip antennas, and metallic thin films,~\cite{Hofmann2007PlasmonicCathodoluminescence, Barnard2011ImagingCathodoluminescence,Bashevoy2007HyperspectralResolution} and to measure the statistics of the emitted light.\cite{Meuret2015PhotonCathodoluminescence, Meuret2017PhotonWells, Meuret2019ComplementaryMicroscope, Meuret2018NanoscaleImaging, Bourrellier2016BrightH-BN} Finally, CL measurements can be employed to characterize photonic band structures and measure the local density of states in nanostructured metallic, semiconductor and dielectric materials.\cite{Kuttge2009LocalCathodoluminescence, Peng2019ProbingSpectrum, GarciaDeAbajo2008ProbingSpectroscopy, Mignuzzi2018}

\subsubsection{Probing electron-light interactions in ultrafast electron microscopes}
\label{sec:expt-ultrafast}

Several TEM-based CL instruments are similar to their SEM analogues, as shown in Fig.~\ref{fig:setups}(c). For instance, collecting mirrors can be used to couple light out of the TEM vacuum chambers. The presence of two collecting mirrors can even allow the measurement of backward and forward radiation independently in some commercial systems.\cite{Silver2015CathodoluminescentMicroscope, Stoger-Pollach2017TransitionCathodoluminescence, Stoger-Pollach2019FundamentalsSemiconductors, Stoger-Pollach2021CoherentMicroscope} A TEM analogue of the spectrally/spatially-resolved SEM CL setup was also built,\cite{Remez2017SpectralRadiation, Remez2019ObservingEmission} with a low-numerical-aperture objective outside the vacuum chamber and an additional rotation degree of freedom to measure radiation at various angles. An interesting advantage of TEM-based solutions is the availability of EELS which allows the measurement of electron energy loss and gain after interacting with a sample. EELS provides a method to probe near-field electron-light interactions and is complementary to CL measurement techniques.~\cite{Polman2019Electron-beamNanophotonics} Since EELS is directly related to the photonic local density of states,~\cite{GarciaDeAbajo2008ProbingSpectroscopy} tomographic techniques have been demonstrated to reconstruct the full three-dimensional local density of states in nanoparticles.~\cite{Horl2017TomographicNanoparticles}

Several techniques have been developed to add time-resolved measurements to the field of SEM- and TEM-based CL. This field has also been inspired by the techniques developed for ultrafast TEM in the Zewail group.\cite{Barwick2015PhotonicsMicroscopy,Yang2010ScanningMicroscopy} Currently, time-resolved CL involves modifications to the electron beam emitter to generate electron pulses by the use of ultrashort laser pulses or by using fast electrostatic beam blanking to modulate a continuous electron beam. An example of a time-resolved TEM instrument is shown in Fig.~\ref{fig:setups}(d), wherein an ultrafast laser is used to generate short time duration electron pulses from the electron source. The ultrafast laser can also be used to excite or probe the sample as a function of the of arrival of the electron pulse. Alternatively, instruments with beam blanking devices located after the electron source can provide time-resolved measurements, albeit at a lower time resolution than the laser-driven emitters. One example of a beam blanking measurement is found in Ref.\cite{Moerland2016Time-resolvedStates}, where a modified standard SEM with beam blanking electronics was used to produce electron pulses in the 80 to 90~ps duration range. This provided sufficient time resolution to characterize the spontaneous emission decay rate in a cerium-doped yttrium aluminium garnet sample.~\cite{Moerland2016Time-resolvedStates}

\subsubsection{Free-electron analogues}
\label{sec:analogues}

Physical analogues of free electrons are convenient platforms to observe some of the above-mentioned physical phenomena. There are many free-electron analogues in the context of CR, since it is a general wave phenomenon, with analogues in classical and quantum electromagnetics, superfluid hydrodynamics, and classical hydrodynamics.\cite{Carusotto2013} In electromagnetics, CR can also be observed with superluminal polarization waves in SPP,\cite{Genevet2015ControlledMetamaterial} quantum cascade lasers,\cite{Belkin2015}, solitons in optical fibers,~\cite{Yuan2011HighlyFiber, Chang2010HighlyGeneration, Skryabin2003SolitonFibers, Zhang2013EnhancedFiber} and superluminal domain perturbations in rapidly time-modulated systems.\cite{Yablonovitch1989AcceleratingEffect, Sloan2021ControllingPerturbations, Oue2021CerenkovGrating, Dikopoltsev2022LightTime-crystals} Synchrotron-like radiation has also been observed by nonlinear polarization induced in a metasurface.\cite{Henstridge2018SynchrotronPulse}

Most of the physical effects discussed in this Review are typically observed from electron beams generated in electron microscopes, with kinetic energies from few keV to hundreds of keV. Analogues of those effects have been predicted with hot electrons in graphene for CR\cite{Kaminer2016EfficientGraphene} and two-dimensional electrons in a driving field for SPR in the THz regime.\cite{Mikhailov2013Graphene-basedEmitter}

Some of those analogue systems have been utilized as a test bed for novel physical phenomena. In particular, metamaterial-loaded and slot waveguides have been used to directly emulate the propagation of an electron beam.\cite{Chen2011FlippingRadiation,Xi2009ExperimentalMetamaterial,Jing2021PolarizationMetasurfaces,Antipov2007WakefieldWaveguides,Antipov2008ObservationWaveguide} The slot waveguide platform was first used to demonstrate backward Cherenkov radiation,\cite{Xi2009ExperimentalMetamaterial} an intriguing effect which has otherwise not been demonstrated in free-electron experiments. They have also been used to demonstrate polarization control in SPR\cite{Jing2021PolarizationMetasurfaces} and, in combination with helical metastructures, SPR vortex beams.\cite{Zhu2020Free-Electron-DrivenEmitter, Zhang2020OrbitalRadiation} 

\subsection{Controlling free-electron radiation in nanophotonic structures}
\label{sec:control}

\begin{figure*}
\includegraphics[scale = 0.8]{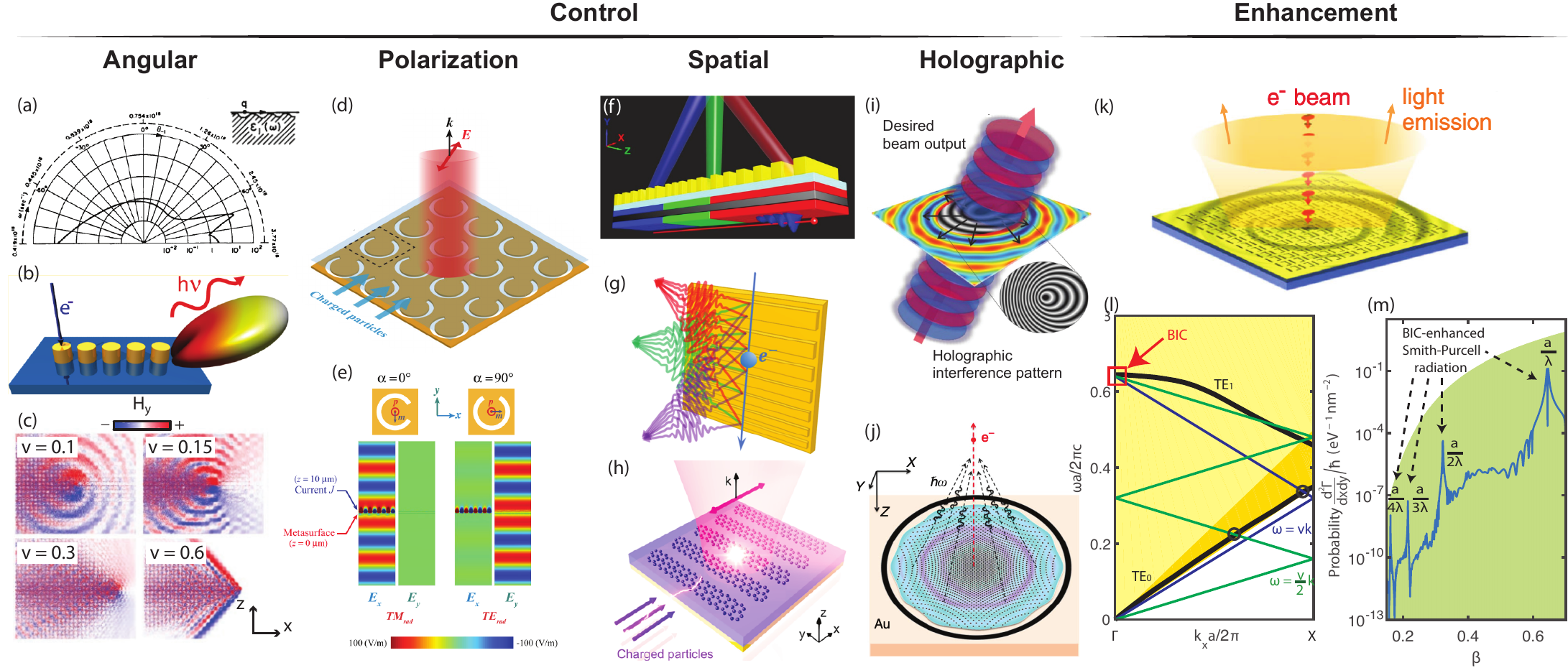}
\caption{\label{fig:controllingsp} \textbf{Controlling and enhancing coherent cathodoluminescence with nanophotonics.} Angular control: (a) Angular SPR distribution in a metallic grating with permittivity profile shown as inset. The sharp peak at
$\theta = 71\text{deg}$ is where the surface-plasmon mode is excited strongly. Reproduced with permission from Ref.\cite{Chuang1984}. (b) Angular control of CL emission in plasmonic Yagi-Uda antennas. Reproduced with permission from Ref.\cite{Coenen2011}. (c) Distribution of radiated magnetic field from free electron propagating in PhC. Inset shows the value of $v$ used in simulation (in units of $c$). Reproduced with permission from Ref.\cite{Luo2003CerenkovCrystals}. Polarization control: (d) SPR polarization control with Babinet metasurfaces. Reproduced with permission from Ref.\cite{Wang2016}. (e) SPR electric-field distributions from C-apertures. Reproduced with permission from Ref.\cite{Wang2016}. Spatial control: (f) Schematic representation of a chirped grating acting as a SPR lens. Reproduced with permission from Ref.\cite{Lai2017GenerationRadiation}. (g) Focused Smith-Purcell emission by a free-electron-driven metalens: free electrons passing in close proximity to a metagrating with a chirped period emit a converging wavefront, with different
wavelengths converging at different positions. Reproduced with permission from Ref.\cite{Karnieli2022CylindricalRadiation}. (h) Graphene metasurfaces offer a playground to control the amplitude, phase, and polarization of SPR. Reproduced with permission from Ref.\cite{Su2019CompleteMetasurfaces}. (i) Schematic of a free-electron holographic light source, a universal approach allowing generation of light with prescribed wavelength, direction, divergence and topological charge via point-excitation of holographic metasurfaces. Reproduced with permission from Ref.~\cite{Li2016}. (j)  Formation of focused broadband transition radiation in interaction of an electron beam with an engineered planar lens. The structure is designed to effectively mimic a porous hemispherical geometry. Reproduced with permission from Ref.~\cite{Talebi2019MergingSources}. Enhancement: (k) Electron-beam-driven collective metamaterial light source concept: highly localized electron beam excitation at the center of a metamaterial array leads, via the strong coupling among metamolecules, to the collective oscillation of many cells and thereby the emission of a free-space light beam. Reproduced with permission from Ref.~\cite{Adamo2012}. (l) The calculated TE band structure (solid black lines) in the $\Gamma$-X direction. The area shaded in light and dark yellow indicates the light cone of air and silica, respectively. The electron lines (blue for velocity $v$, and green for $v/2$) can phase match with either the guided modes (circle) or high-Q resonances near a BIC (red square). Reproduced with permission from Ref.\cite{Yang2018MaximalElectrons}. (m) Probability of free-electron-induced photon emission in a two-dimensional PhC slab as a function of reduced electron velocity. Strong enhancement of SPR matched phase-matched to BIC and analytical limit accounting for material losses (shaded green). Reproduced with permission from Ref.\cite{Yang2018MaximalElectrons}.}
\end{figure*}

\subsubsection{Angular and spectral control}
\label{sec:control-angular}

The spectral-angular distribution of coherent CL effects is, to first order, embedded into their dispersion relation (e.g., Eq.~(\ref{eq:spdisp}) for SPR). When free electrons emit in a nanophotonic medium, certain spectral and angular components of the radiative fields can be selectively enhanced. The possibility of selective enhancement is evident in periodic structures, where the emitted energy is proportional to the overlap between the electron trajectory and Fourier components of the photonic modes (as in Eqs.~(\ref{eq:sp1d},~\ref{eq:sp3d})). The shaping of the spectral-angular distribution via photonic engineering has been the focus of many recent works in coherent CL. Certain notable results are shown in Fig.~\ref{fig:controllingsp}.

It was first proposed by Van den Berg that even perfect reflectors with sinusoidal corrugations could significantly alter the angular distribution of SPR.\cite{VandenBerg1973} Controlling the emission direction can also be achieved by exciting plasmonic resonances in periodic metallic gratings with a free electron.\cite{Chuang1984} SPR spectral-angular shaping has also been proposed and demonstrated with engineered grating profiles,~\cite{Remez2017SpectralRadiation} resulting in multi-peaked spectra, and aperiodic gratings.\cite{Saavedra2016Smith-PurcellArrays}

Localized electron beam excitation can also be leveraged to achieve directional emission in single resonators\cite{Coenen2014DirectionalScatterer} (selectively exciting and interfering multipolar modes) and plasmonic Yagi antennas excited by free electrons.\cite{Coenen2011}

The control of radiative angular distribution has also been proposed in CR in PhC\cite{Luo2003CerenkovCrystals} and resonant TR in multi-layer structures.\cite{Lin2018ControllingRadiation,Lin2021ADetectors} It was demonstrated for TR in elliptical plasmonic bull's eye targets.\cite{Coenen2019Energy-MomentumAntennas} 

\subsubsection{Polarization control}
\label{sec:control-polarization}

Coherent CL effects are strongly linearly polarized, with limited tunability of the polarization angle. For instance, the polarization angle in SPR is set by the propagation direction of the electron beam. The control of the linear polarization angle and the generation of spin angular momentum in CL has therefore received recent interest, with some notable works shown in Fig.~\ref{fig:controllingsp}(d-e). 

In CR, small components of circular polarization can be acquired when considering spin-polarized electron beams.\cite{Kaminer2016QuantumMomentum, Peshkov2020Spin-polarizationElectrons} With unpolarized electron beams, metasurfaces on waveguides have been proposed as platforms to generate circularly polarized CR.\cite{Li2020CircularWaveguides} Because of its potential applications in spectroscopy, the generation of circularly-polarized TR light in the mm-wave regime was realized by interfering its forward and backward components\cite{Shibata2001GenerationRegion}

In particular, SPR is highly polarized along the direction of the electron beam (a feature which has been known since its original discovery\cite{Smith1953}). Recent work using cross-coupled electric and magnetic dipoles in THz Babinet metasurfaces suggested a path to steer the angle of linear polarization.\cite{Wang2016, Yang2018ManipulatingMetasurfaces, Liu2017TerahertzEfficiency} In principle, graphene metasurfaces can also be utilized to generate circular polarization states in the THz regime.\cite{Su2019CompleteMetasurfaces} More recently, an experimental demonstration was provided by exciting cross-polarized resonances in a PhC.\cite{Yang2021ObservationResonances} The generation and control of SPR with orbital angular momentum (vortex beams) could also be achieved with holographic gratings.\cite{Wang2020VortexGrating} Full control of the optical angular momentum (spin and angular) in SPR could also find applications in on-chip spectroscopy, but its realization has remained elusive thus far. 

\subsubsection{Spatial control}
\label{sec:control-spatial}

As coherent CL relies on the excitation of photonic eigenmodes, phase relationships between different wave vector components are constrained, which prevents the control of the radiation far-field profile. With the promise of on-chip electron-driven light sources, there has been a growing interest in manipulating the far-field spatial distribution of emitted radiation in CL. If successful, this effect could be leveraged to realize integrated sources and collimators into a single component, with some notable works shown in Fig.~\ref{fig:controllingsp}(f-h).

The design of a CR lens was first proposed by adjusting the boundary of the Cherenkov target based on ray optics considerations to concentrate light into a focal spot.\cite{Galyamin2014DielectricRadiation, Galyamin2018FocusingTheory} Given the well-defined spectral-angular relation in SPR, a natural design for a SPR concentrator is a chirped grating (with decreasing pitch along the beam trajectory). That way, the emission angle is tuned along the direction of propagation or, equivalently, a thin-lens-like phase modulation is imparted to light generated via SPR. This concept was first proposed theoretically and demonstrated numerically for concentrators working at single wavelengths,\cite{Remez2017SpectralRadiation, Lai2017GenerationRadiation} and exhibiting strong chromatic aberrations. Alternatively, graphene metasurfaces can also generate converging SPR in the THz regime.\cite{Su2019CompleteMetasurfaces} Recently, signatures of SPR lensing have been reported\cite{Karnieli2022CylindricalRadiation} using a chirped grating design. 

The most general type of far-field wavefront engineering, holography, has also been proposed using tailored nanophotonic structures (see Fig.~\ref{fig:controllingsp}(i, j)).~\cite{Li2016} This method relies on the controlled interference of transition radiation, generated by a focus electron beam, with an interference holographic mask. This enables the generation of light with prescribed wavelength, direction, divergence and topological charge via point-excitation of CL holography in plasmonic~\cite{Li2016,Clarke2018Direction-divisionSources, Schilder2020Phase-ResolvedHolography} and dielectric~\cite{Clarke2018All-dielectricSources} metasurfaces. Free-electron holographic light sources offer a universal approach to generate light with prescribed wavelength, direction, divergence and topological charge via point-excitation of holographic metasurfaces with an electron beam. Lastly, inspired by transformation optics, several nanophotonic structure designs have been demonstrated to realize broadband focusing of transition radiation,~\cite{Talebi2019MergingSources} vortex light beams,~\cite{VanNielen2020ElectronsSieves} and more generally structured light from free electrons.~\cite{Christopher2020Electron-drivenMicroscopes} Such structures have been proposed as platforms to measure time-energy correlation functions in electron microscopes, paving the way towards attosecond electron-based spectroscopy techniques.~\cite{Talebi2016SpectralMicroscopes}.

\subsection{Enhancing free-electron-light interactions in nanophotonics}
\label{sec:enhance}

\subsubsection{Coherent cathodoluminescence}
\label{sec:enhance-coherent}


The existence of fundamental bounds on CL, as presented in section~\ref{sec:theory-bounds}, begs the following question: could one enhance coherent CL with nanophotonics to achieve emission efficiencies approaching such bounds? 


The possibility of resonant enhancement in electron-light interaction is highlighted in the phase-matching relation Eq.~(\ref{eq:phase-matching}). Emission into phase-matched photonic modes is selectively enhanced by adjusting the electron velocity. This concept has been proposed to enhance SPR by coupling electrons to photonic\cite{Yamaguti2002PhotonicRadiation} and plasmonic resonances,\cite{Chuang1984} and CR in PhCs.\cite{Luo2003CerenkovCrystals} Resonant enhancement of the electron-light interaction is also observed in an increase in PINEM signal when exciting photonic resonances.\cite{Wang2020CoherentCavity, Kfir2020ControllingModes}

Specifically, one can design a resonance mode of interest in metamaterials and excite it with a beam of free electrons. In particular, localized free-electron-beam excitation can create a low-divergence spatially coherent free-space light beam that bears similarity with laser light through coherent collective oscillation of an ensemble of coupled metamolecules.\cite{Adamo2012}

An interesting feature of the general bound from Eq.~(\ref{eq:bound}) is its apparent divergence in the limit of small losses. Recently, the use of BICs~\cite{Hsu2016} was theoretically proposed as a new
mechanism for enhanced SPR: coupling of
electrons with BICs.\cite{Yang2018MaximalElectrons} Such photonic modes have the extreme quality factors
of guided modes but are, crucially, embedded in the radiation continuum, with no resulting SPR into
the far field. Fig.~\ref{fig:controllingsp}(j) shows that by tuning the electron velocity (here, a sheet electron beam translationally invariant along the $y$ direction), one can achieve strong emission enhancements (such as in CL into a guided mode), while keeping the radiative coupling into a continuum resonance. This enhancement technique also requires a large modal overlap between the BIC and the electron near field (see Fig.~\ref{fig:controllingsp}(i)).  This enhancement mechanism is in line with the upper limits from Eq.~(\ref{eq:bound}), since the enhancement is limited by the material's non-zero losses at the emission wavelengths. Nevertheless, it has been theoretically shown that BIC-enhanced coupling enables
the radiation intensity to closely approach this upper limit at several resonant velocities. In the presence of an absorptive channel, the maximum enhancement occurs at a small offset from the BIC where the $Q$-matching condition is satisfied.

Photonics can also provide CL enhancement via band structure engineering in periodic structures. Specifically, the perspective of enhancing SPR and CR was first proposed in two-dimensional periodic PhCs, where the intersection of the electron plane with photonic band structures can be manipulated.\cite{Kremers2009TheorySpectrum} In particular, it was predicted that bands with vanishing transverse components of the group velocity would display strong emission enhancement; this has been suggested recently as a platform to realize full phase-matching of point electrons with photonic modes (with a continuum of phase-matched transverse modes). Recent experimental demonstration of resonant enhancement from flatbands also shows their potential in enhancing electron-light interactions.\cite{Yang2021ObservationResonances}

\subsubsection{Incoherent cathodoluminescence and scintillation}
\label{sec:enhance-incoherent}

\begin{figure*}
\includegraphics[scale=0.85]{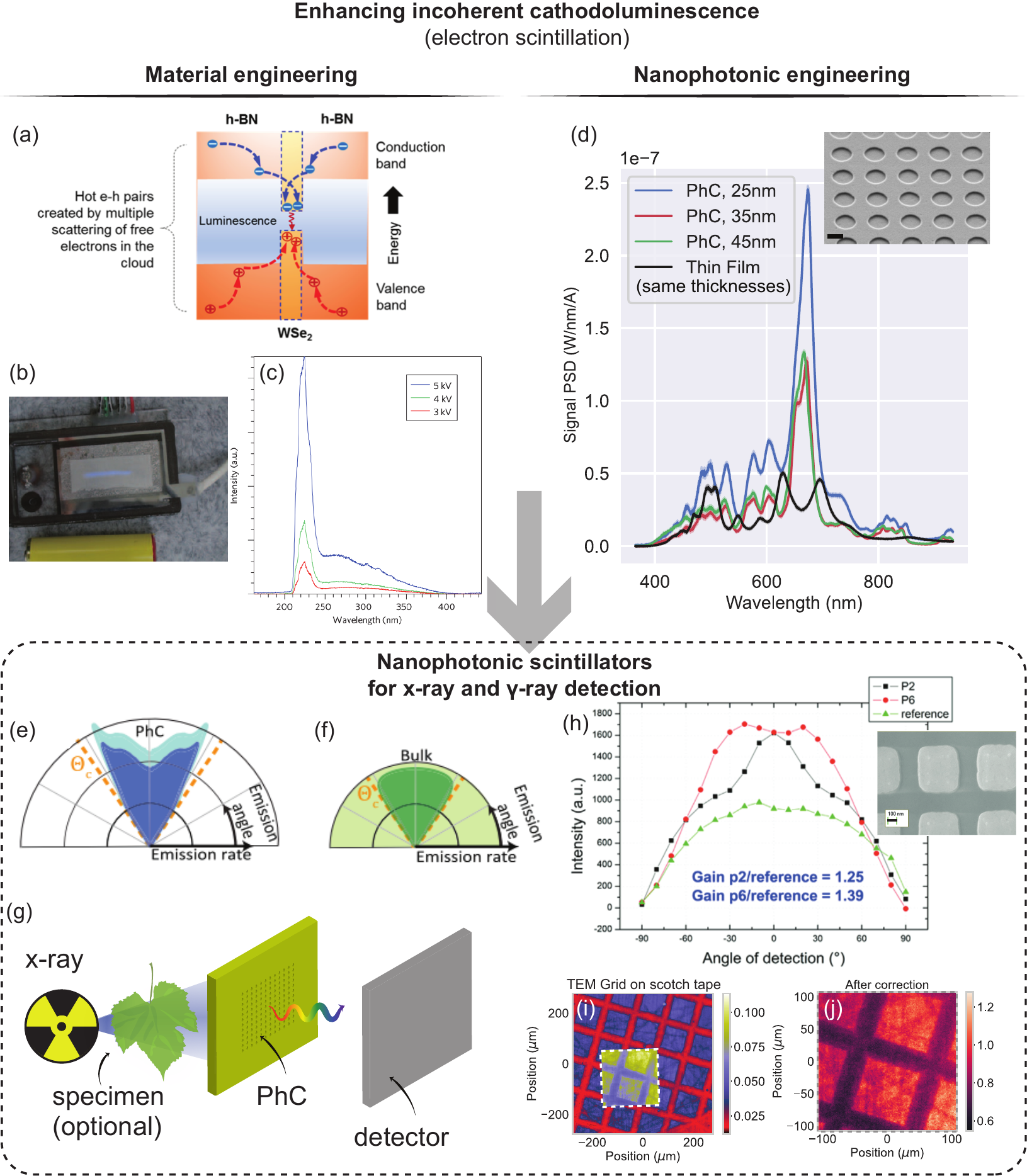}
\caption{\label{fig:scintillation} \textbf{From shaping incoherent cathodoluminescence (ICL) to nanophotonic scintillators.} (a) Nearly monochromatic cathodoluminescence can be observed from monolayers of chalcogenide semiconductor in a van der Waals heterostructure when it is sandwiched between layers of boron nitride with a higher energy gap. Reproduced with permission from Ref.\cite{Zheng2017GiantSemiconductors}. (b) Photograph of a prototype far-ultraviolet plane-emission device in operation. Reproduced with permission from Ref.\cite{Watanabe2009Far-ultravioletNitride} (c) Output spectra show a dominant peak at 225 nm. This broad far-ultraviolet band is a result of superposition of excitonic bands from 215 to 227 nm. The weak tail from 250 to 400 nm is a result of impurities and defect states. Reproduced with permission from Ref.\cite{Watanabe2009Far-ultravioletNitride}. (d) ICL signal from self-trapped holes defect in silicon-on-insulator wafer for various PhCs and thin film geometries. Inset shows SEM image of PhC with etch depth 35~nm (scale bar 200~nm). Reproduced with permission from Ref.\cite{Roques-Carmes2022ANanophotonics}. (e) Angular distribution from a multilayer scintillator. Reproduced with permission from Ref.\cite{Kurman2020Photonic-CrystalDetection}. (f) Angular distribution from bulk scintillator. Reproduced with permission from Ref.\cite{Kurman2020Photonic-CrystalDetection}. (g) Schematic of PhC scintillators for x-ray applications. (h) Measured scintillation angular distribution from scintillators with dielectric antireflection coatings. Inset: scanning electron micrograph of sample P6 (scale bar 100~nm). Samples "P2" and "P6" refer to two different designs of antireflection coatings on scintillators. Reproduced with permission from Ref.\cite{Knapitsch2015ReviewScintillators}. (i) X-ray scan of a TEM grid glued to tape using a PhC scintillator on YAG:Ce. The yellow bright area corresponds to the PhC pattern. Reproduced with permission from Ref.\cite{Roques-Carmes2022ANanophotonics}. (j) Flat-field corrected image from (i). Reproduced with permission from Ref.\cite{Roques-Carmes2022ANanophotonics}.}
\end{figure*}

The process of light emission from fluctuating current sources in samples pumped by high-energy particles is called ICL (for free-electron pumps) or scintillation (for x- and $\gamma$-ray pumps). From the multi-physics picture illustrated in Fig.~\ref{fig:overview}(e), there appears to be at least two ways to enhance it: (1) control of the available emitting states in the electronic band structure (material engineering) and; (2) control of the nanophotonic environment (nanophotonic engineering). 

\paragraph{Material engineering.} 
\label{sec:enhance-incoherent-material}

ICL can be enhanced by engineering the electronic band structure of emitting materials, and specifically available defect states or band gap to make bright emitters. Such techniques are also ubiquitous in the context of brighter and faster x-ray scintillator development, which rely on the development of new materials\cite{Gektin2017InorganicSystems} (typically, with large atomic numbers, high density, and doped with rare earth elements to emit strongly in the UV-NIR). 

In the context of ICL, two-dimensional materials and van der Waals heterostructures have received particular attention, given their bright emission properties and exceptional nanophotonic properties.\cite{Basov2016PolaritonsMaterials, Caldwell2019PhotonicsNitride} Specifically, bright UV emission from hexagonal boron nitride has been demonstrated, with emission patterns reminiscent of a lasing behavior.\cite{Watanabe2004} Such substrates were then integrated into handheld devices to demonstrate their applicability to sanitization\cite{Watanabe2009Far-ultravioletNitride} (see Fig.~\ref{fig:scintillation}(c,~d)). Those achievements have been enabled by the control of hexagonal boron nitride's excitonic properties, by growing crystals under high pressure and temperature. Recent work utilized ICL as a probe of the material's properties down to the few-layer regime.\cite{Elias2019DirectNitride}

In semiconductors, the emission process is similar to LEDs, and the electron-hole pair recombination probability limits the emission efficiency. In van der Waals heterostructures, engineering of the excitonic properties of an ICL emitter can be realized, for instance by "sandwiching" a transition metal dichalcogenide monolayers between two layers of hexagonal boron nitride, resulting in several orders of magnitude ICL enhancement.\cite{Zheng2017GiantSemiconductors} 

Another common method for enhancing ICL and scintillation is by fabricating semiconductor quantum dots whose small dimensions cause quantum confinement of the excited charge carriers, resulting in larger dipoles and enhanced emission rates.~\cite{Chen2018All-inorganicScintillators}

\paragraph{Nanophotonic engineering.}
\label{sec:enhance-incoherent-nanophotonic}

The other avenue to enhance ICL and scintillation is to engineer the nanophotonic environment in which the fluctuating current sources from Eq.~(\ref{eq:randcurr}) radiate. The possibility of nanophotonic enhancement can be explained by Eqs.~(\ref{eq:scint}, \ref{eq:scint-reciprocal}). 

Scintillation and ICL being in essence spontaneous emission, they can be enhanced with nanophotonics in two distinct ways: (1) via direct enhancement of the rate of spontaneous emission by shaping the density of optical states \cite{Kurman2020Photonic-CrystalDetection, Purcell1946SpontaneousFrquencies} (as evident from the eigenmode formulation in Eq.~(\ref{eq:scint})); or (2) through improved light extraction from bulk scintillators (as evident from the far-field formulation in Eq.~(\ref{eq:scint-reciprocal})). The prospect of enhancing scintillation through the local density of states, as well as the prospect of large scintillation enhancements, by either mechanism, had remained unrealized until recently.\cite{Roques-Carmes2022ANanophotonics} 

A recent demonstration of enhanced ICL from self-trapped hole defects in silica is shown in Fig.~\ref{fig:scintillation}(e), where the emitted ICL from a defect transition emitting around $\sim700$~nm is enhanced and spectrally-shaped. This is enabled by the presence of several high-$Q$ photonic resonances due to the PhC patterning at the surface of the scintillator. In the same work, ICL is used as a spectroscopic probe of the nonlinear microscopic dynamics accounting for competing emission processes. Given the generality of Eqs.~(\ref{eq:scint}, \ref{eq:scint-reciprocal}), we note that similar techniques can be used to enhance emission from a stochastic ensemble of emitters, which could find applications beyond ICL and scintillation: in thermal radiation,\cite{Greffet2002CoherentSources,Greffet2018LightLaw, Overvig2021ThermalInteractions, Iyer2020UnidirectionalMetasurfaces} LEDs,\cite{Liu2018Light-EmittingRadiation} and photoluminescence.\cite{Liu2018Light-EmittingRadiation}

\subsubsection{Nanophotonic scintillators for x- and $\gamma$-ray detection}
\label{sec:enhance-incoherent-scintillators}

Our approach highlights the common physical origin of ICL (from free electrons) and scintillation from x- and $\gamma$-rays. In a broader context scintillators are used for applications across medical imaging, non-destructive testing, and night vision technologies.\cite{Gektin2017InorganicSystems} This analogy holds especially for low-energy x-rays whose penetration depths are comparable to electrons in SEM and TEM. 

The possibility of nanophotonic scintillation enhancement for x- and $\gamma$-ray application was recently revived by the availability of large-scale deposition and patterning techniques compatible with state-of-the-art scintillators in x-ray and nuclear imaging. Early work demonstrated enhanced light extraction provided by a PhC coating atop a bulk scintillator \cite{Liu2018EnhancedEmbossing,Knapitsch2012ResultsScintillators,Knapitsch2015ReviewScintillators,Zhu2015EnhancedLithography,Pignalosa2012GiantNanostructures,Liu2021ImprovedCrystals,Ouyang2020EnhancedCrystals} (see Fig.~\ref{fig:scintillation}(i)). It was then realized that x-ray scintillation could be enhanced via local photonic density of state enhancement in multi-layered structure, a convenient platform to also shape the scintillation angular emission profile.\cite{Kurman2020Photonic-CrystalDetection} Taking a step forward, a general framework for scintillation in nanophotonics was proposed,~\cite{Roques-Carmes2022ANanophotonics} with experimental demonstrations of both enhancement mechanisms in two distinct platforms (ICL and x-ray scintillation). The perspective of using such nanophotonic scintillators for x-ray imaging was demonstrated in the same work (see Fig.~\ref{fig:scintillation}(j,k)). 

\section{Future perspectives}
\label{sec:future}

In this section, we outline several research directions, which we believe will significantly flourish in the next few years. In particular, we highlight the use of nanophotonics in enabling these novel applications of (quantum) electron-light interactions.

\begin{figure*}
\includegraphics[scale = 0.72]{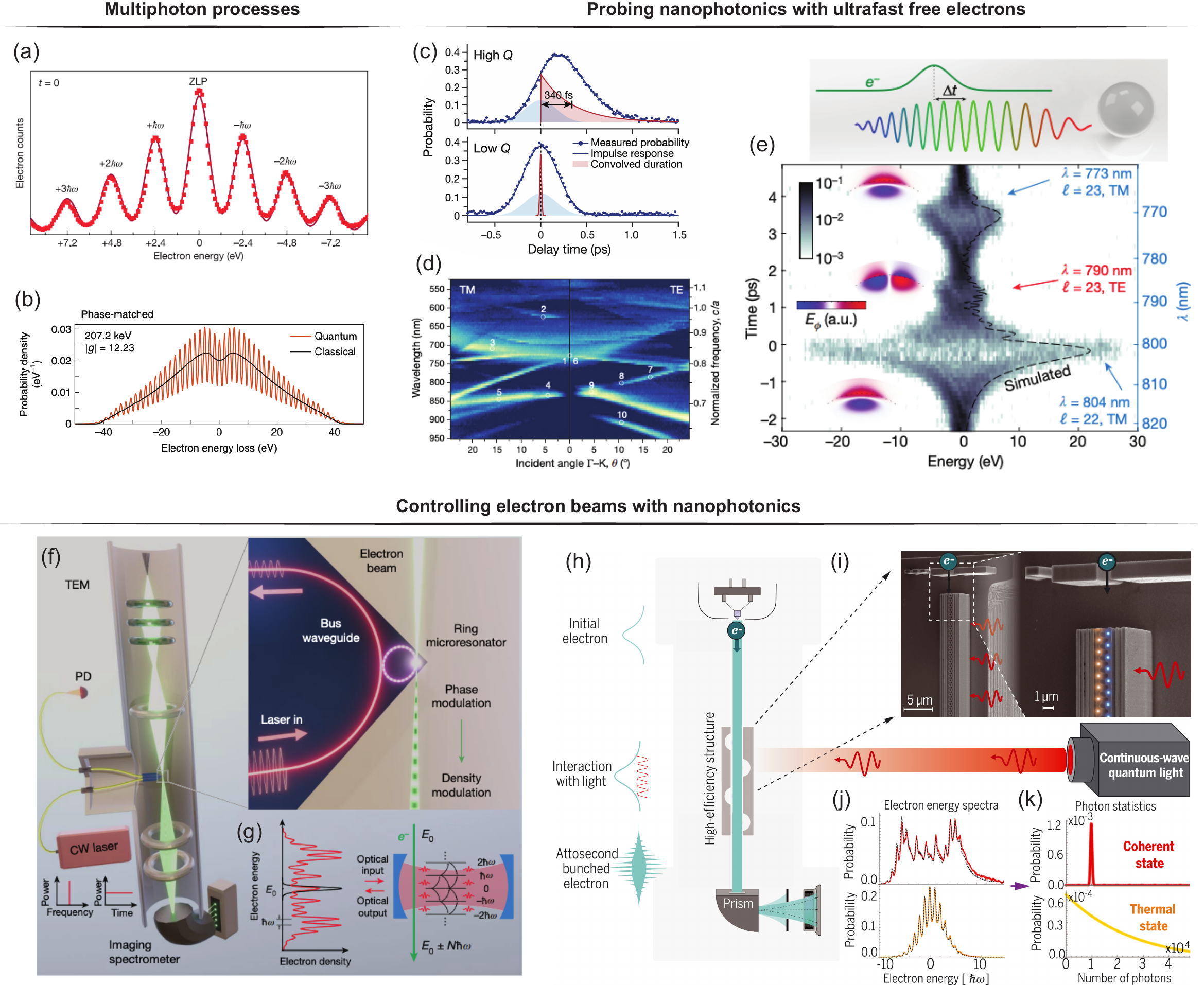}
\caption{\label{fig:quantum} \textbf{Quantum and ultrafast free-electron-light interactions.} (a) Electron energy loss spectra (EELS) of carbon nanotubes irradiated with an intense fs laser pulse at two different delay times. The energy spectrum at coincidence of the two pulses ($t= 0$~fs) displays the zero-loss peak (ZLP) and multiple quanta of photon absorption/emission. Reproduced with permission from Ref.\cite{Barwick2009Photon-inducedMicroscopy}. (b) Electron energy loss spectrum for a perfectly phase-matched interaction between a free-electron and a photonic near field along a prism. Reproduced with permission from Ref.\cite{Dahan2020ResonantWavefunction}. (c) Experimental
(dots) and simulated (solid curve) probabilities of the electron–cavity-photon interaction as a function of delay time, enabling the measurement of photonic $Q$-factors. Reproduced with permission from Ref.\cite{Wang2020CoherentCavity}. (d) Band structure measured by scanning over incident laser angles and wavelengths. Each data point in the map is a separate EELS measurement of the electron–light interaction at zero delay time. Reproduced with permission from Ref.\cite{Wang2020CoherentCavity}. (e) Probing multiple nanophotonic resonances with EELS and a chirped pulse. The colour maps
present the relevant azimuthal component of the electric field. Reproduced with permission from Ref.\cite{Kfir2020ControllingModes}. (f) Rendering of the experimental setup to achieve continuous-wave modulation of an electron beam. Inset: magnified interaction region with the electron beam passing a microresonator. CW, continuous wave; PD, photodiode. Reproduced with permission from Ref.\cite{Henke2021IntegratedModulation}. (g) Left: during interaction with the laser-driven cavity mode, the initially narrow electron spectrum (black) develops discrete sidebands at integer multiples of the photon energy (red). Right: in a cavity quantum electrodynamics depiction, the cavity photons induce transitions between the free-electron energy ladder states. Reproduced with permission from Ref.\cite{Henke2021IntegratedModulation}. (h) CW modulation of electron wave functions in transmission electron microscopy. Reproduced with permission from Ref.\cite{Dahan2021ImprintingElectrons}. (i) Highly efficient electron-light interaction facilitated by an inverse-designed silicon-photonic nanostructure (scanning electron microscope image), consisting of a Bragg mirror and a periodic channel that achieves quasi-phase-matching of electron and quantum light. Reproduced with permission from Ref.\cite{Dahan2021ImprintingElectrons}. (j) The electron energy spectrum after the interaction with two types of light states: coherent and thermal. Reproduced with permission from Ref.\cite{Dahan2021ImprintingElectrons}. (k) The corresponding photon statistics reconstructed from the measured spectra. Reproduced with permission from  Ref.\cite{Dahan2021ImprintingElectrons}.}
\end{figure*}

\begin{figure}[ht!]
\includegraphics[scale = 0.8]{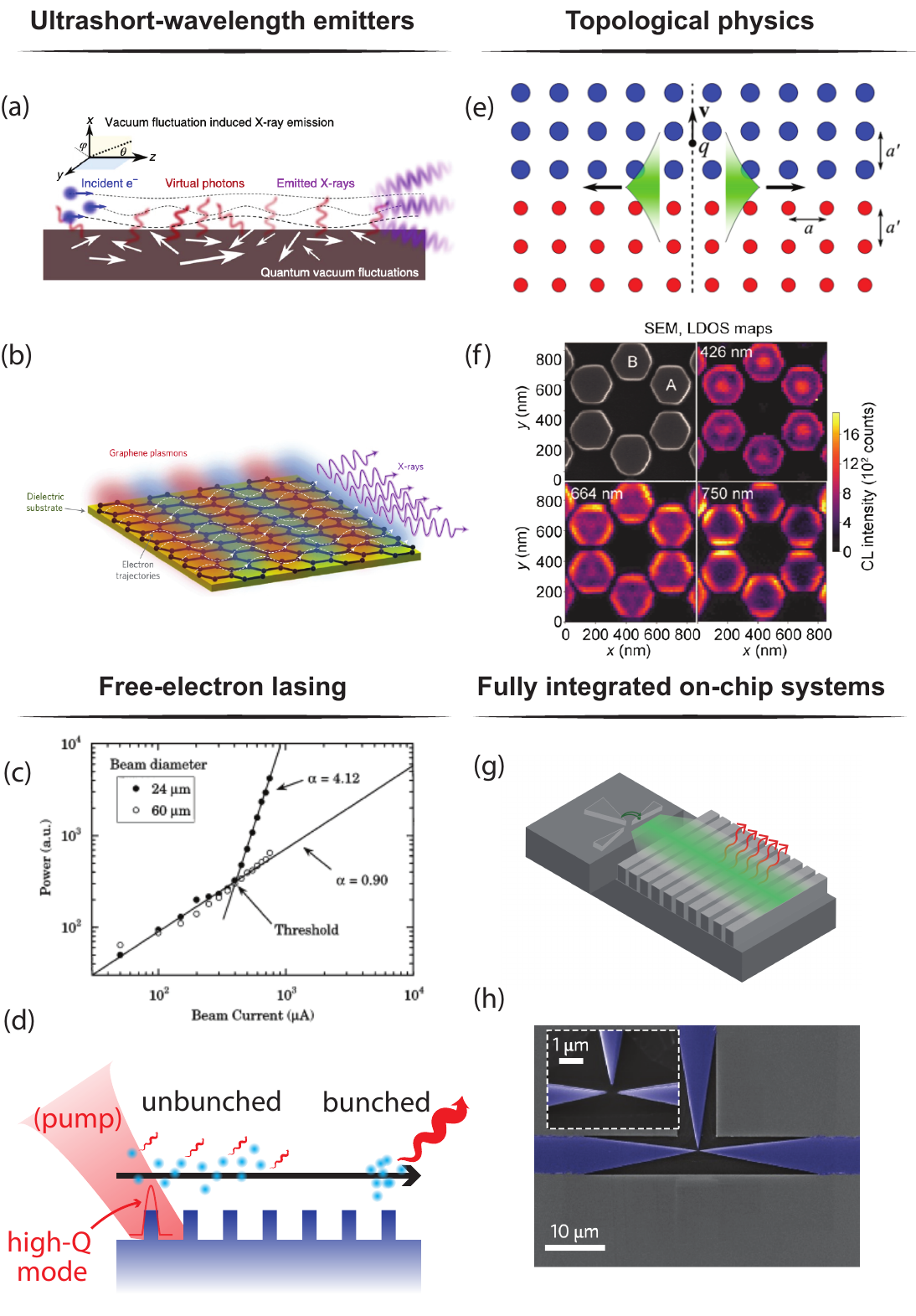}
\caption{\label{fig:futureprospects} \textbf{Future prospects in free-electron radiation.} (a) Schematic illustrating a beam of electrons travelling in the near field of a nanophotonic structure. The vacuum fields lead to random modulations of the trajectory with a non-zero variance, resulting in significant x-ray emission. Reproduced with permission from Ref.\cite{Rivera2019LightForces}. (b) Compton scattering with graphene plasmons: free electrons (dotted white lines) interact with the graphene plasmon field (glowing red and blue bars) to produce short-wavelength output radiation. Reproduced with permission from Ref.\cite{Wong2015}. (c) Detected power vs. electron beam current in an experiment demonstrating superradiant SPR in the 300-900~$\mu$m range. Reproduced with permission from Ref.\cite{Urata1998SuperradiantEmission}. (d) Proposal for free-electron lasing in nanophotonic structures. Unbunched electron beam interacting with a high-$Q$ photonic mode can self-bunch. The use of an optional pump can enable the observation of stimulated emission from free electrons. (e) TR between two PhCs, resulting in the excitation of edge and bulk modes. Reproduced with permission from Ref.\cite{Yu2019TransitionCharge}. (f) Spatially-resolved cathodoluminescence spectroscopy of topological modes in a hexagonal photonic lattice of silicon Mie resonators. Reproduced with permission from Ref.\cite{Peng2019ProbingSpectrum}. (g) Fully-integrated on-chip Smith-Purcell silicon radiator, consisting of a silicon field emitter array and a silicon nanograting. Reproduced with permission from Ref.~\cite{Roques-Carmes2019TowardsSources}. (h) Planar molybdenum electron emitter used for on-chip Cherenkov emission from hyperbolic metamaterials. Reproduced with permission from Ref.~\cite{Liu2017IntegratedThreshold}.}
\end{figure}

\subsection{Quantum effects}


Much of the content of this Review can be understood in purely classical terms: in terms of solutions to the classical Maxwell's wave equation in a dielectric medium, with a source $\mathbf{J}(\mathbf{r}, \omega)$ (a moving electron). Even in the case of ICL (scintillation), the radiation may be understood in terms of radiation from classically non-deterministic current sources, although the correlations of the current sources may be dictated by the quantum statistics of the radiating currents (e.g., Bose-Einstein or Fermi-Dirac statistics). Nowadays, there is a great deal of effort in theoretically predicting, as well as experimentally isolating, genuinely quantum mechanical effects in free-electron radiation, which may not be described in terms of the classical wave equation with a source. In this section, we give a brief outline of the different types of quantum effects that can emerge in free-electron radiation. For more discussion of such effects (with different emphasis), we refer the reader to Ref. \cite{Rivera2020LightmatterQuasiparticles, GarciaDeAbajo2021OpticalOpportunities}. 

\subsubsection{Recoil effects} 

As discussed in section~\ref{sec:theory}, much of the spectral and angular properties of coherent cathodoluminescence may be understood from the phase-matching relation Eq.~(\ref{eq:phase-matching}). This phase-matching, while naturally following from the classical Maxwell's equations, can also be understood in quantum mechanical terms. Taking the emitted photon to have a four-momentum $k = \hbar(\omega,\mathbf{k})$ and the electron to have its own (initial $(i)$ / final $(f)$) four-momentum $p_{i,f} = (E_{i,f},\mathbf{p}_{i,f})$, with $E$ the electron energy and $\mathbf{p}$ the electron momentum - one finds that the phase-matching condition is nothing other than an approximate solution to the energy-momentum conservation condition $p_i = p_f + k$: under the approximation that the electron energy and momentum change only weakly due to the emission. For almost all cases in electron-photon interactions in nanophotonics, this approximation is very accurate, since the typical energy of the photon ($\approx$1~eV), is quite small compared to that of the electron (in excess of $\approx$1~keV). The main regimes in which the recoil approximation may break down are: low electron energies,\cite{Tsesses2017LightNanostructures, Talebi2016SchrodingerEffect, Huang2022QuantumCrystals} high photon energies,\cite{Wong2015, Rivera2019LightForces, Rosolen2018Metasurface-basedSource} and near velocity thresholds\cite{Kaminer2016QuantumMomentum} (between which the emission is kinematically forbidden or allowed). 
All three of these regimes have been subject of theoretical investigation: as of this writing, clear quantum recoil effects in coherent CL have not been observed (for SPR, the region where such quantum effects should arise is shown in Fig.~\ref{fig:spoverview}). 

\subsubsection{Quantum interference and waveshaping effects} 
Another way by which quantum effects can alter light emission by electrons is through electron waveshaping. One might imagine that controlling the wavefunction of the emitting electron will control the emitted radiation in much the same way that controlling the shape of the current density in Maxwell's equations controls the radiation. The situation in quantum mechanics is more subtle, as the wavefunction of the electron does not represent a smeared out charge density. If it did, one could imagine that simply expanding or compressing the electron wavefunction transverse to its motion would change its radiation (e.g., by Cherenkov or Smith-Purcell emission): tunable transverse coherent size has shown that the intensity does not depend on the electron wavefunction.~\cite{Remez2019ObservingEmission} In fact, the transverse size had no influence on the emitted radiation, as first predicted for CR.~\cite{Kaminer2016QuantumMomentum} Despite this, the shape of the electron wavefunction \emph{can} alter the radiation, under specific circumstances: namely, if the emitted photon has enough symmetries broken. For example, in a recent experiment involving EELS, it was found that the electron energy-loss spectrum could be modified by shaping the electron wavefunction to have its symmetry be compatible with different localized (plasmonic) modes of a structure.\cite{Vanacore2017} Namely, the electron could lose energy primarily through a dipolar mode of the structure (of one energy) when the electron wavefunction had a certain symmetry, and lose energy primarily through a quadrupolar mode (of a different energy) when the electron had a different symmetry. This shows the influence of the electron wavefunction on plasmon emission.

The key constraint that enables the dependence on the wavefunction is that the measurement system performs post-selection on the electron (by conditioning the measurement on the electron state). Without electron post-selection, the radiation intensity is independent on the electron wavefunction.~\cite{Kaminer2016QuantumMomentum, GarciaDeAbajo2021OpticalOpportunities, DiGiulio2021ModulationLight} Unlike with the radiation intensity, other properties of the radiation such as coherence can be modified by shaping the electron wavefunction. This was first predicted in Ref.~\cite{Karnieli2021TheParticle}, showing that the pulse duration of CR depends on the electron wavefunction. Building on this concept, electron wavefunction shaping was proposed as a method to control the quantum states of free-electron radiation.~\cite{BenHayun2021ShapingElectrons, Wong2022TailoringWavepackets}

Recent theoretical work has also shown how the emission of high-energy photons such as x-rays can be controlled through quantum interference effects.\cite{Wong2021ControlWavepackets} In that work, it was shown how quantum interference of amplitudes could lead to strong control over phenomena such as bremsstrahlung radiation, where an electron emits radiation after scattering off of an external (Coulomb) potential. In particular, two different initial electron states can scatter into the same final electron and emitted photon states, leading to quantum interference. Such interference can only occur when the external potential breaks sufficient symmetries to enable both transitions to conserve energy and momentum simultaneously. Beyond this, the effect of the electron wavefunction on the entanglement structure of the emitted radiation has been explored theoretically in other examples of coherent CL (e.g., CR).\cite{Kaminer2016QuantumMomentum} Ref.~\cite{Kfir2021OpticalElectrons} also showed the existence of phase correlations between the emitted CL field and the electron-modulating laser in the quantum-optical regime.

\subsubsection{High-order effects in quantum electrodynamics} A third class of important effects, which are now a rich and vibrant field on their own, is related to strong-field interactions, in which an electron can interact with a strong external field, leading to absorption or stimulated versions of coherent CL (see Fig.~\ref{fig:quantum}). For the types of fields that are now typical in such experiments, an electron can emit and absorb multiple (even hundreds) of quanta from the external field, leading to a large number of quantized peaks in the electron energy loss/gain spectrum of the post-interaction electron.\cite{Dahan2020ResonantWavefunction} Since the first observations of this effect in 2009,\cite{Barwick2009Photon-inducedMicroscopy} which forms the basis for PINEM, this effect has been observed and extended in many guises and has been exploited for imaging electromagnetic fields in materials with high spatial (nanometer) resolution and high temporal (sub-picosecond) resolution.\cite{Park2010Photon-inducedExperimental,Piazza2015SimultaneousNear-field,Feist2015QuantumMicroscope, Priebe2017AttosecondMicroscopy, Morimoto2018DiffractionTrains, Vanacore2018AttosecondFields, Vanacore2017, Vanacore2019UltrafastFields, Wang2020CoherentCavity,Dahan2020ResonantWavefunction} Of particular note is that the electron energy spectrum after the interaction is sensitive to the quantum statistics of the field,~\cite{Gorlach2020UltrafastElectrons} as recently demonstrated in Ref.~\cite{Dahan2021ImprintingElectrons} to distinguish between thermal and coherent light. Such effects form the basis for the extension of quantum optics to free electrons. High-order processes have also been proposed in CR from heavy ions, resulting in modified emission angles and suppression of radiation pathways close to the threshold.~\cite{Roques-Carmes2018NonperturbativeEffect} More recently, quantum regimes of interactions between free electrons and photons were observed at much lower energies in a SEM.~\cite{Shiloh2022Quantum-CoherentMicroscope}

\subsection{Ultrashort-wavelength emitters}
\label{sec:future-ultrashort}

Free electrons are ideal platforms to generate high-energy radiation, since they carry large kinetic energies (in the keV to MeV range in most electron microscopes) that can be converted into electromagnetic radiation. Among the effects we have discussed in this Review, SPR displays the most evident wavelength tunability, via the incident electron velocity and the structure periodicity. In Fig.~\ref{fig:spoverview}, we highlighted recent work towards SPR from low-energy (sub-keV) electrons and short-wavelength photons. Specifically, there has been excitement around achieving tunable SPR in the UV.\cite{GarciadeAbajo1999} There has been renewed interest in using SPR to generate UV light, with recent experimental demonstrations down to a wavelength of 230~nm. The perspective of generating UV light with SPR is especially promising for sanitization applications.\cite{GarciaDeAbajo2020BackSpaces} 

Generation of much shorter wavelength radiation has been reported in crystals, whose structure is naturally "modulated" at \aa ngstr\"{o}m scales. Recently, such effects found renewed interest in van der Waals heterostructures pumped by free electrons emitting soft x-ray light,\cite{Shentcis2020TunableMaterials} with enhanced functionality demonstrated in Ref.~\cite{Huang2022EnhancedStructures}. The ability to control coherent CL with engineered nanostructures is especially exciting for short-wavelength radiation generation, given the unavailability of efficient focusing optics in this wavelength range. Therefore, the ability to concentrate coherent CL has received specific attention recently, with proposals in the near-UV\cite{Remez2017SpectralRadiation} and x-ray regime.\cite{shi2022free, Balanov2021TemporalInteraction}

Pumping of wide-bandgap materials with free electrons is another way to generate bright UV light. In particular, hexagonal boron nitride was proposed as a bright UV emitter, exhibiting lasing behavior,\cite{Watanabe2004} and the possibility of being integrated in compact devices.\cite{Watanabe2009Far-ultravioletNitride} The existence of a large catalogue of wide-bandgap semiconductors,\cite{Takahashi2007WideDevices} in addition to the availability of nanophotonic enhancement and shaping,\cite{Roques-Carmes2022ANanophotonics} and the theoretical possibility of large conversion efficiencies,\cite{Klein1968BandgapSemiconductors} makes UV ICL a particularly promising research direction. 

Other avenues, such as scattering of electrons with graphene SPPs\cite{Wong2015, Rosolen2018Metasurface-basedSource, Pizzi2020GrapheneSources} (see Fig.~\ref{fig:futureprospects}(a)) and virtual vacuum photons\cite{Rivera2019LightForces} (see Fig.~\ref{fig:futureprospects}(b)) have been proposed to generate table-top x-ray sources pumped by free electrons.  

\subsection{Nanophotonic dielectric laser accelerators}
\label{sec:dla}
The time-reversal process of free-electron emission generation is particle acceleration. Similarly, nanophotonic development leads to new designs, device miniaturization, and performance improvement in particle accelerators.~\cite{Plettner2008ProposedUndulator, England2014, Breuer2013, Peralta2013, Shiloh2021ElectronAcceleration} 

A channel for electron propagation usually has to be defined for accelerators; this limits the current applied to the device. To address this issue, a multi-channel accelerator structure was proposed,~\cite{Zhao2020DesignAccelerator} which naturally transforms the seminal dual-grating accelerating structure into PhCs, thus allowing device designs via 2D and 3D band structure engineering. The use of photonic flatband proposed for enhancing SPR~\cite{Yang2021ObservationResonances} can be used in reverse to aid in optical localization and enhancement of electron acceleration without the need for electron channels, thus allowing more electrons to interact with accelerator devices.

Experimentally, an inverse-designed (see section below), waveguide-integrated, microscale dielectric laser accelerator was fabricated and tested, showing acceleration gradients around 30~MeV/m in a recent demonstration.~\cite{Sapra2020On-chipAccelerator} 
Meanwhile, electron phase space control was realized via introducing phase discontinuity (similar to a lattice dislocation) in a high-aspect-ratio, silicon-based, integrated structure,~\cite{Shiloh2021ElectronAcceleration} indicating the possibility for particle acceleration of confined beams to the MeV range with minimal particle loss over extended distances. Quantized peaks in electron energy spectra have also been observed in dielectric laser accelerators, revealing the quantum nature of such devices.\cite{Adiv2021QuantumAccelerators}

\subsection{Inverse-designed free-electron radiators and accelerators}
\label{sec:future-inversedesign}
Computational inverse design has now penetrated every single field of nanophotonics.~\cite{Molesky2018InverseNanophotonics, Jensen2011TopologyNano-photonics} Free-electron radiation will be no exception. In general, inverse design enables the automatic discovery of nanophotonics structures that are locally optimal for a pre-defined objective function. Among the various inverse-design techniques available, topology optimization is one of the most popular, given the many degrees of freedom it can investigate, while remaining computationally tractable, thanks to efficient gradient calculation methods (such as the adjoint method\cite{Molesky2018InverseNanophotonics} and automatic differentiation techniques\cite{Jin2020InversePropulsion}). Topology optimization enables the inverse design "pixel by pixel" of various nanophotonic components, such as metasurfaces,\cite{Lin2018Topology-OptimizedMetaoptics} multiplexers,\cite{Piggott2015InverseDemultiplexer} and nonlinear frequency converters,\cite{Lin2016Cavity-enhancedOptimization} while taking into account fabrication constraints.\cite{Hammond2021PhotonicConstraints} 

Topology optimization has recently been utilized to discover optimal nanophotonic structures as free-electron radiators, nanophotonic scintillators,\cite{Roques-Carmes2022ANanophotonics} radiators and accelerators.\cite{Haeusler2022BoostingDesign, Sapra2020On-chipAccelerator} Several features of free-electron radiation make it amenable to inverse design. First, electron-light interaction lends itself to closed-form expressions describing radiation and acceleration processes (such as the total emitted power in a range of angles, frequencies, and polarizations; or the particle acceleration gradient). Those expressions are enabled by the modeling of electron-light interactions mediated by the current sources from Eqs.~(\ref{eq:detcurr}, \ref{eq:randcurr}). Second, numerical methods are available in the time- and frequency- domains to calculate those expressions and their respective gradients efficiently, either via the adjoint method, or via automatic differentiation algorithms. This enables the implementation of efficient gradient-based discovery of complex nanophotonic structures interacting efficiently with free electrons. We also expect that recent efforts in inverse design for free-electron physics will bolster interest in fully-differentiable multi-physics pipelines, integrating electron trajectories and energy loss, microscopic dynamics, and nanophotonics.\cite{Roques-Carmes2022ANanophotonics}

\subsection{Free-electron lasing in nanophotonic structures and other exotic effects}
\label{sec:future-lasing}

As discussed earlier in this Review, the various coherent mechanisms by which a free electron can emit light (e.g., CR, SPR, TR) can be thought of as elementary processes of spontaneous emission in which an electron transitions to a state of lower kinetic energy while emitting a quantized mode of Maxwell's equations (a photonic quasiparticle \cite{Rivera2020LightmatterQuasiparticles}), whose properties (e.g., energy, momentum, polarization, and other quantum numbers) strongly control the observed properties of the emission. From this ``quantum'' perspective of free-electron radiation, the general principles of quantum mechanics determine that an electron should be able to absorb photonic quasiparticles (leading to inverse processes, such as inverse CR\cite{Fontana1998AEffect}) and also \emph{stimulatedly emit} photonic quasiparticles - leading to amplification and lasing phenomena.\cite{Schachter1989Smith-PurcellLaser} 

Although we have used quantum arguments to infer the possibility of stimulated emission and free-electron lasing, much of the known phenomena can be described classically. In fact, the "stimulated emission" of light by a beam of classical electrons simply corresponds to net deceleration of a beam of electrons in an applied electromagnetic field: by energy conservation, the lost energy of the electrons goes into the field (into photons). Meanwhile, "absorption" of photons by the electrons corresponds to net acceleration. 

Net stimulated emission by a beam of electrons forms the basis for a number of developed and developing technologies. In the microwave domain, these effects form the basis for established high-power microwave sources such as traveling wave tubes and klystrons used in space communication systems.\cite{Pierce1950TravelingWaveTubes, Xu2020ReviewRadiation} In traveling wave tubes and klystrons, such ``lasing'' is largely based on SPR. In the x-ray domain, amplification has also been achieved based on the collective interactions of a dense beam of electrons with a spatially periodic magnetic field. In such systems, the electrons can amplify their own spontaneous emission, when the density of the electron beam is sufficiently high. 

As of this writing, free-electron lasing based on coherent CL in nanophotonic systems is yet to be observed, despite reported signatures of superradiance at THz frequencies,\cite{Urata1998SuperradiantEmission} shown in Fig.~\ref{fig:futureprospects}(c). Such observations represent not only a logical step in the development of compact and integrated free electron radiation sources, but may also grant some practical advantages. Compared to CL, the coherence of the lasing field should be quite high (especially in architectures based on an optical feedback cavity,\cite{Andrews2004GainLaser, Pellegrini2016TheLasers} such as a PhC resonance,\cite{Rivera:20,Kumar2006AnalysisLasers} as shown in Fig.~\ref{fig:futureprospects}(d)). 

Lastly, other unconventional effects in free-electron radiation, such as backward CR, remain unobserved as of this writing. Originally predicted in left-handed materials,~\cite{Veselago1968ElectrodynamicsPermeabilities, Chen2011FlippingRadiation, Luo2003CerenkovCrystals, Wu2007Left-handedRadiation, Matloob2004CerenkovMedium} and then in PhCs,~\cite{Luo2003CerenkovCrystals} this effect has not been observed yet with free electrons (despite observations using free-electron analogues~\cite{Xi2009ExperimentalMetamaterial}). The observation of this effect in nanophotonics with free electrons could open a way to novel high-energy particle detector designs.


\subsection{Topological effects in free-electron radiation}
\label{sec:future-topological}
Topological photonics \cite{Lu2014TopologicalPhotonics,Ozawa2019TopologicalPhotonics} has undergone rapid development over the past decade. So far, optical excitations have been mostly employed for these studies.
In previous sections, we have shown that photonic topological features, such as BICs~\cite{Yang2018MaximalElectrons, Song2018CherenkovLasers} and flatbands,~\cite{Yang2021ObservationResonances} can be used for boosting radiation generation from free electrons.
In the following, we outline other opportunities at the intersection between free-electron nanophotonics and topological photonics.

Several recent advances show that free electrons can serve as probes for photonic topology. In Ref.~\cite{Peng2019ProbingSpectrum}, the local density of states and the band structure of a topological PhC were mapped out via CL techniques, as shown in Fig.~\ref{fig:futureprospects}(f). 
In Ref.\cite{Yu2019TransitionCharge}, highly relativistic electron beams were used to excite the domain-wall edge states of a quantum-spin-Hall-like metamaterial, where localization to the edge was identified (see Fig.~\ref{fig:futureprospects}(e)). 
In Ref.\cite{Yu2020UltrafastMetamaterials}, free electrons were proposed for resolving the optically-driven phase transition of a graphene-dielectric metamaterial.
Besides, it is worth mentioning that photonic topological properties can also be potentially adapted, in turn, for diagnosing the location, energy, and duration of charged particle beams.

Furthermore, free electrons can serve as pumps for photonic topology. It has been shown that free electron beams can drive nonreciprocity in metallic waveguides, therefore giving rise to unidirectionally-propagating modes.~\cite{Fallah2021NonreciprocalBeams} Because the free electron velocity can be substantially higher than that in solids, the bandwidth of the resulting unidirectionality could be dramatically increased. 

Some of the free-electron analogues mentioned earlier in this Review could be adequate platforms to demonstrate physical effects which require hard-to-realize photonic topological states in the visible regime such as chiral edge states\cite{Wang2009ObservationStates} and Weyl points.\cite{Lu2015ExperimentalPoints}

\subsection{Non-equilibrium emission processes with free electrons}
\label{sec:future-noneq}

The recent development of a general framework to predict non-equilibrium radiation phenomena such as scintillation (or ICL) reveals the sensitive dependence of the scintillation spectrum on the non-equilibrium distribution of the excited electrons generated by the high-energy electron beam.~\cite{Roques-Carmes2022ANanophotonics} In particular, the radiated photons can inherit the non-equilibrium statistical properties of the excited electrons and holes generated in the scintillation process. Consider an interesting example which has yet to be observed in scintillation: when electrons and holes in semiconductors are produced by a high-energy pump, the electrons and holes quickly relax into a non-equilibrium quasi-steady state which can be described by quasi-Fermi levels for the electrons and holes which are different. In such a case, the radiated photons must (by detailed balance) experience an effective chemical potential dictated by the difference between the electron and hole chemical potentials.\cite{Wurfel1982TheRadiation,Greffet2018LightLaw} This is of course in complete contrast to thermal radiation, in which the photons have no chemical potential. More broadly, by controlling the energy of the high-energy electron pump, different non-equilibrium distributions of electrons and holes can be conceivably created, shaping the quantum statistical properties of the radiated photons. The ability to achieve this would represent a new way to control the quantum optical properties of photons. Such effects could be measured through second-order intensity correlation ($g^{(2)}$) measurements, as applied to ICL,\cite{Meuret2015PhotonCathodoluminescence, Bourrellier2016BrightH-BN} where it was shown that for low enough electron-currents, the radiation could show signatures of antibunching, indicating single-photon emission, or signatures of bunching, indicating super-Poissonian statistics, depending on the material.

\subsection{Fully integrated on-chip systems}
\label{sec:onchip}

The use of compact nanoscale structures to convert energy from free electrons into radiation has spurred interest into realizing fully on-chip systems. This perspective is enabled by the realization of on-chip electron emitters, such as field emitter arrays.~\cite{Guerrera2016, Guerrera2016a, Temple1999, Spindt1991Field-EmitterMicroelectronics}. Such devices can realize focused electron beams, in continuous wave or pulsed regimes, with relatively low kinetic energies (from few tens of keV to few hundreds of eV). 

It was first realized that Smith-Purcell and Cherenkov systems could be fully integrated on chip \cite{Tang1996, Ishizuka2001, Ishizuka2000SmithPurcellCathode, Neo2006Smith-PurcellEmitter}, thereby enabling integrated, tunable light emitters. One of the additional advantages of fully integrated systems is the ability to precisely align the electron emitter with nanophotonics structures, thereby optimizing free-electron emission. One such device is shown in Fig.~\ref{fig:futureprospects}(g), with a recent design of electron emitter shown in Fig.~\ref{fig:futureprospects}(h). Recent work focused on the realization of all-silicon Smith-Purcell emitters~\cite{Roques-Carmes2019TowardsSources} and on-chip systems emitting CR from hyperbolic metamaterials.~\cite{Liu2017IntegratedThreshold} The realization of fully integrated electron accelerator devices has also been the focus of much research recently.~\cite{Sapra2020On-chipAccelerator, Shiloh2021ElectronAcceleration, Niedermayer2021DesignChip,Niedermayer2017DesigningChip,Black2019Laser-DrivenMicrostructures}

With the development of vacuum packaging techniques~\cite{Lemoine2009}, in addition to
high-voltage DC-DC converters~\cite{Behnam2008} for lab-on-chip applications, such integrated systems could be realized in a scalable way and be compatible with CMOS fabrication processes.

\section{Conclusion}
\label{sec:conclusion}

In conclusion, we have provided an extensive review of the fundamental physics and applications of free-electron interactions with nanophotonic structures. We first proposed a general classical framework to model cathodoluminescence as the excitation of photonic eigenmodes by the free-electron current density. This excitation can be coherent, as is the case for all coherent CL effects, or partially incoherent, as is the case for ICL and scintillation. This framework enabled us to re-derive fundamental formula in free-electron physics, and to provide a unified method to shape and enhance CL. It also enabled us to derive fundamental bounds on electromagnetic radiation from free electrons in arbitrary nanophotonic environments. 

Those methods have been confirmed and observed thanks to the development of experimental methods to observe and characterize light emission from free electrons in nanophotonic structures. Specifically, we reviewed the development of CL and EELS measurement techniques in SEM and (ultrafast) TEM. The several experimental methods we reviewed have enabled the measurement of all optical properties of interest: spectral, angular, time, spatial, polarization distributions, and quantum correlations. They can also be complemented by measurement of the electron properties with EELS. 

Finally, we shared our enthusiasm for a few avenues in free-electron radiation enabled by nanophotonics, namely the emerging fields of ultrafast, topological, and quantum effects in electron microscopy, the design of ultrashort-wavelength emitters pumped by free electrons, miniaturized particle accelerators, novel inverse-design techniques applied to free-electron emitters, and the aspiration to reach stimulated emission and lasing by free electrons in nanophotonics. While the realization of some of those effects might still be a few years ahead of us, we believe that the strong interest from our community has already demonstrated the potential of nanophotonics in controlling and enhancing CL. Future developments in the field might unlock some of the most promising prospects in free-electron physics, such as the realization of widely tunable integrated light sources from x-rays to mm-wave, particle accelerators, and high-energy particle detectors. They should find direct applications across particle detection and acceleration, electron microscopy, dosimetry, water sanitization, medical imaging and therapy.

\section{Authors' contribution}
C.~R.-C. wrote the manuscript with inputs from S.~E.~K., Y.~Y., N.~R., and P.~D.~K. All authors read and reviewed the final manuscript.

\section{Data availability}
The data that support the findings of this study are available
within the article and upon request for Fig.~\ref{fig:spoverview}.

\section{Acknowledgments}
The authors would like to thank Nikolay I. Zheludev, Kevin MacDonald, and Liang Jie Wong for their helpful comments on the Review. This material is based on work supported in part by the US Army Research Laboratory and the US Army Research Office through the Institute for Soldier Nanotechnologies under contract W911NF-18-2-0048. This material is also in part based on work supported by the Air Force Office of Scientific Research under awards FA9550-20-1-0115 and FA9550-21-1-0299.

\appendix

\section{Appendix: Unified theory of coherent cathodoluminescence}
In this Appendix, we provide more details on the derivation of emitted power densities for coherent CL. We then re-derive emblematic formulas of the field for CR, TR, and excitation of SPPs.

We derive our theory in the context of classical electrodynamics, with the analogy of a moving free electron as a time-dependent current source. We consider the case of a moving electron of charge $q$ propagating along the linear trajectory defined by the speed vector $\mathbf{v}=v\hat{r}_{\parallel}$, which is described by the current distribution of Eqs.~(\ref{eq:detcurr}, \ref{eq:detcurr_freq}).

This expression of the current in Fourier domain already highlights the existence of preferred field wavevectors to which the electron beam will couple to, which is the essence of the phase-matching condition shown in Eq.~(\ref{eq:phase-matching}). The nanophotonic environment is described by the Green's function expansion shown in Eq.~(\ref{eq:green}).

The total energy radiated by a dipole can be calculated as an integral in frequency domain:\cite{Kremers2009TheorySpectrum}
\begin{equation}
    U = -\frac{1}{\pi} \text{Re} \left( \int d\omega \int d\mathbf{r} ~  \mathbf{J}^*(\mathbf{r}, \omega) \mathbf{E}(\mathbf{r}, \omega) \right).
\end{equation}

We then get the master equation shown in Eq.~(\ref{eq:genform}), which describes all coherent CL effects considered in this Review. In the next sections, we provide more details on its application the cases of CR, TR, and excitation of SPPs. 

\subsection{Cherenkov radiation}
\label{sec:appendix-cherenkov}

CR in its simplest embodiment occurs in a homogeneous dielectric environment and consists in the spontaneous emission of plane waves by a charged particle (see Fig.~\ref{fig:nomenclature}(a)). Its description entails eigenmodes as plane waves of the form $\mathbf{F}_k(\mathbf{r}, \omega) = \frac{e^{i\mathbf{k}\cdot \mathbf{r}}}{\sqrt{V\tilde{\epsilon}}}\hat{\epsilon}_\mathbf{k}$, in a uniform medium of refractive index $n=\sqrt{\epsilon}$ with polarization vector $\hat{\epsilon}_\mathbf{k}$ orthogonal to $\mathbf{k}$, and $\tilde{\epsilon} = (2\omega_k)^{-1}\frac{d}{d\omega}(\omega^2\epsilon(\omega))|_{\omega = \omega_k}$. The dispersion relation is $\omega_k = c|\mathbf{k}|/n$. We also make the following adjustment $\sum_m \rightarrow \int V \frac{d\mathbf{k}}{(2\pi)^3}$. Injecting this expression into Eq.~(\ref{eq:genform}), we get the Frank-Tamm formula\cite{Frank1937CoherentMatter} for the spectral density per unit propagation length:
\begin{equation}
    \label{eq:franktamm}
    \frac{dU}{d\omega dl} = \frac{\mu_0 q^2}{4\pi} \omega  \sin^2\theta ~ \Theta(\beta n - 1),
\end{equation}
with $\Theta$ the Heaviside function ($\Theta(x) = 1$ if $x>0$, otherwise $\Theta(x) = 0$). Here, the Cherenkov angle is defined such as $\cos\theta = (\beta n)^{-1}$. The Heaviside function defines the well-known Cherenkov threshold $\beta n > 1$, which is equivalent to satisfying the phase-matching condition from Eq.~(\ref{eq:phase-matching}) in a bulk medium.  

\subsection{Transition radiation}
\label{sec:appendix-tr}
To describe transition radiation within the same framework, we consider the simplified case of a charge impinging at normal incidence on a perfect conductor (see Fig.~\ref{fig:nomenclature}(c)). We resort to the introduction of an image charge with opposite charge and velocity $(-q, -\mathbf{v})$. The current distribution $\mathbf{J}(\mathbf{r},\omega)$ is modified accordingly and its emission in free space is considered. The current distribution is given by:
\begin{align}
    \label{eq:detcurr_tr}
    \mathbf{J}(\mathbf{r}, t) & = q \mathbf{v} \left( \delta(\mathbf{r}-\mathbf{v}t) + \delta(\mathbf{r}+\mathbf{v}t) \right) ~ \text{for} ~ t<0\\ 
    \mathbf{J}(\mathbf{r}, t) & = 0 ~ \text{for} ~ t>0,
\end{align}
and its Fourier transform
\begin{equation}
    \label{eq:detcurr_freq_tr}
    \mathbf{J}(\mathbf{r}, \omega) = q \hat{r}_\parallel \delta(\mathbf{r}_{\perp}) e^{-i\omega\sigma(r_\parallel)\frac{r_\parallel}{v}},
\end{equation}
with $\sigma$ the sign function. 
Utilizing the free space mode expansion as in the case of Cherenkov radiation, we get the following spectral distribution, first derived by Ginzburg and Tamm:\cite{Ginzburg1940QuantumMedium}
\begin{equation}
    \label{eq:tr}
    \frac{dU}{d\omega d\Omega} = \frac{\mu_0 q^2 \beta^2 c}{4\pi^3} \frac{\sin^2 \theta}{\left(1-\beta^2\cos^2\theta\right)^2}.
\end{equation}

\subsection{Coherent excitation of SPPs by free electrons}
\label{sec:appendix-spp}

Excitation of SPPs is often considered in tandem with TR, since it typically occurs at the interface between two media supporting SPP. This phenomenon is observed even in the absence of corrugation at the surface because the electron point nature enables coupling directly to the SPP modes. Our formalism captures this effect by considering emission into the SPP modes rather than free space modes as before.

For simplicity, we consider an electron flying parallel to a planar interface at $r_\perp = 0$ (reduced to a 2D problem). We consider the simplified case of a lossless interface between a dielectric (relative permittivity $\epsilon_1>0$) and a metal (relative permittivity $\epsilon_2<0$). The $p$-polarized modes take the form (in the $(r_\parallel, r_\perp)$ basis):
\begin{align}
\mathbf{F}_{1,\mathbf{k}}(\mathbf{r})&\propto
    \begin{pmatrix}
        1 \\
        -\sqrt{-\frac{\epsilon_1}{\epsilon_2}} \\
    \end{pmatrix}
    e^{-|k_{1,\perp}|r_\perp} e^{-ik_{\parallel}r_\parallel} \\
\mathbf{F}_{2,\mathbf{k}}(\mathbf{r})&\propto
    \begin{pmatrix}
        1 \\
        -\sqrt{-i\frac{\epsilon_1}{\epsilon_2}} \\
    \end{pmatrix}
    e^{-|k_{2,\perp}|r_\perp} e^{-ik_{\parallel}r_\parallel}.
\end{align}
And the following dispersion relation: 
\begin{equation}
    k_\parallel^2 = \frac{\epsilon_1\epsilon_2}{\epsilon_1+\epsilon_2}\frac{\omega^2}{c^2}, ~\text{and}~ k_{\perp,i}^2 = \frac{\epsilon_i^2}{\epsilon_1+\epsilon_2}\frac{\omega^2}{c^2}.
\end{equation}

We obtained the following expression for the energy emitted by the free electron per unit length:
\begin{equation}
    \label{eq:spp}
    \frac{dU}{d\omega dl} \propto \frac{q^2\mu_0c}{2\pi^2} \sqrt{\frac{-\epsilon_1^2}{\epsilon_1+\epsilon_2}}e^{-2|k_\perp|r_\perp},
\end{equation}
with the phase-matching condition:
\begin{equation}
    \omega = v k_\parallel = \beta \sqrt{\frac{\epsilon_1 \epsilon_2}{\epsilon_1 + \epsilon_2}} \omega,
\end{equation}
which can be solved graphically by finding the intersection of the SPP's dispersion and the electron line $\omega = v k_\parallel$. Such an intersection always exists for the lower branch (corresponding to the excitation of a SPP).

\section*{References}
\label{sec:references}
\bibliography{references}

\end{document}